\documentclass[aip,amsmath,amssymb,seceqn,10pt]{revtex4-1}
\usepackage{graphicx}
\usepackage{graphics}
\usepackage{epsfig}
\usepackage{dcolumn}
\usepackage{bm}
\usepackage{makeidx}
\usepackage{subfigure}
\usepackage{color}
\usepackage{longtable}
\usepackage{framed}
\usepackage{hyperref}

\begin{document}

\title{Topological antiferromagnetic spintronics: Part of a collection of reviews on antiferromagnetic spintronics}

\author{Libor~\v{S}mejkal}
\affiliation{Institute of Physics, Academy of Sciences of the Czech Republic, Cukrovarnicka 10, 162 00 Praha 6 Czech Republic}
\affiliation{Institut fur Physik, Johannes Gutenberg Universitat Mainz, D-55099 Mainz, Germany}
\affiliation{Faculty of Mathematics and Physics, Charles University in Prague,
Ke Karlovu 3, 121 16 Prague 2, Czech Republic}
\author{Yuriy~Mokrousov}
\affiliation{Peter Gr\"{u}nberg Institut and Institute for Advanced Simulation, Forschungszentrum J\"{u}lich and JARA, 52425 J\"{u}lich, Germany}
\author{Binghai~Yan}
\affiliation{Department of Condensed Matter Physics,
Weizmann Institute of Science, 7610001 Rehovot, Israel}
\author{Allan~H.~MacDonald}
\affiliation{Department of Physics, University of Texas at Austin, Austin, Texas 78712-0264, USA}

\date{\today}

\begin{abstract}
The recent demonstrations of electrical manipulation and detection of antiferromagnetic spins have opened up
a chapter in the spintronics story.  In this article, we review the emerging research field
that is exploring synergies between antiferromagnetic spintronics and topological structures in real and momentum
 space.  Active topics include proposals to realize 
Majorana fermions in an antiferromagnetic topological superconductors, to control topological protection of 
Dirac points by manipulating antiferromagnetic order parameters, and to exploit the anomalous and 
topological Hall effects of zero net moment antiferromagnets. 
We explain the basic physics concepts behind these proposals, and 
discuss potential applications of topological antiferromagnetic spintronics.  
\end{abstract}

\maketitle

Topologically protected states of matter are unusually robust because they cannot be destroyed by
small changes in system parameters.  This feature of topological states has suggested 
an appealing strategy to achieve useful quantum computation \cite{Sarma2015,Beenakker2016}. 
In spintronics, topological states provide for strong  
spin-momentum locking \cite{Hasan2010}, high charge-current to spin-current
conversion efficiency\cite{Fan2016b,Wang2016d,Soumyanarayanan2016}, large electron mobilities and 
long spin diffusion lengths \cite{Pesin2012b,Liang2014}, strong magnetoresistance \cite{Liang2014}, 
and efficient spin-filtering \cite{Wu2014}.  
Materials exhibiting topologically protected Dirac or Weyl quasiparticles in their momentum-space
bands, and those exhibiting topologically non-trivial real-space spin textures \cite{Fert2013,Soumyanarayanan2016}, 
have both inspired new energy-efficient 
spintronics concepts\cite{Pesin2012b,Fert2013,Burkov2016,Felser2016,Fan2016b,Smejkal2017}.
 
In a topological insulator (TI), time reversal symmetry enforces Dirac quasiparticle surface states with spin-momentum locking (see Fig.\ref{Fig1}(a)) and protection against backscattering \cite{Hasan2010}. 
The much higher efficiency of magnetization switching by a current induced spin-orbit torque (SOT) in a 
TI/magnetically doped TI (MTI) heterostructure \cite{Fan2014a,Fan2016b} 
than in a heavy-metal/ferromagnet (FM) bilayer
is thought to be associated with spin-momentum locking, and is a 
paradigmatic example of the potential 
seen for topological materials in spintronics.
(A full microscopic understanding of the underlying 
current-spin conversion mechanism is however still absent \cite{Han2017}.)
Progress in understanding and exploiting topological insulators in spintronics has so far been 
limited by unintentional bulk doping in TIs, and by the decreased stability of TI surface states at 
elevated temperatures \cite{Han2017}.  
The practical utility of the topologically enhanced  SOT  
has also been limited by the cryogenic temperatures at which known MTIs order\cite{Fan2014a,Fan2016b}  
\footnote{A recent report \cite{Katmis2016} of interfacial ferromagnetism
persisting to room temperature at a insulating ferromagnet (EuS) /TI heterostructure 
is promising in this respect.}. 

A substantial rise in the critical temperature of a magnetic topological insulator 
(by a factor of $~$3 to $~$90\,K)
due to proximity coupling to adjacent antiferromagnet (AF) 
has recently been demonstrated in a heterostructure consisting of the metallic antiferromagnet CrSb sandwiched 
between two MTIs \cite{He2016}.  Increased SOT efficiency at heterojunctions between
TI's and ferrimagnetic CoTb alloys containing antiferromagnetically coupled Co and 
Tb sublattices \cite{Finley2016} has also been reported. 
The later effect persists to room temperature, but with decreased efficiency 
enhancement \cite{Han2017}
at higher temperatures.  Research on using antiferromagnetism to achieve a role
for topological materials in spintronics is however still at an early stage and many ideas
have so far only been addressed theoretically.    
The practical advantages of TIs over heavy-metal systems for spin-orbit torques, 
for example, are not yet established \cite{Han2017}.  The forms of magnetism 
so far incorporated in magnetic TIs remain fragile because they are of 
interfacial\cite{Katmis2016,He2016} or dilute-moment character\cite{Fan2014a}.
Other new ideas, beyond simply making topological insulators magnetic,
 are emerging at a rapid pace.
In this article we review topological 
antiferromagnetic spintronics, the emerging field
that is exploring the interplay between
transport, topological properties in either momentum space or real space, and 
antiferromagnetic order.

\section{Topological Insulators in Antiferromagnets and Majorana fermions}

The roots of topological antiferromagnetic spintronics can be traced
to studies of layered AFs of the SrMnBi$_{2}$ type, which were thought to feature 
quasi 2D massive Dirac quasiparticles around the Fermi level\cite{Park2011a,Wang2011e}. 
These were associated with the observation of enhanced mobilities, similarly to those in graphene. 
Masuda {\it et al.}\cite{Masuda2016} have demonstrated manipulation of the Dirac quasiparticle 
current and the quantum Hall effect in a EuMnBi$_{2}$ AF by an applied strong magnetic field,
with the effect of the field mediated by Eu sublattices. 

\begin{figure*}[h!]
  \includegraphics[width=0.7\textwidth]{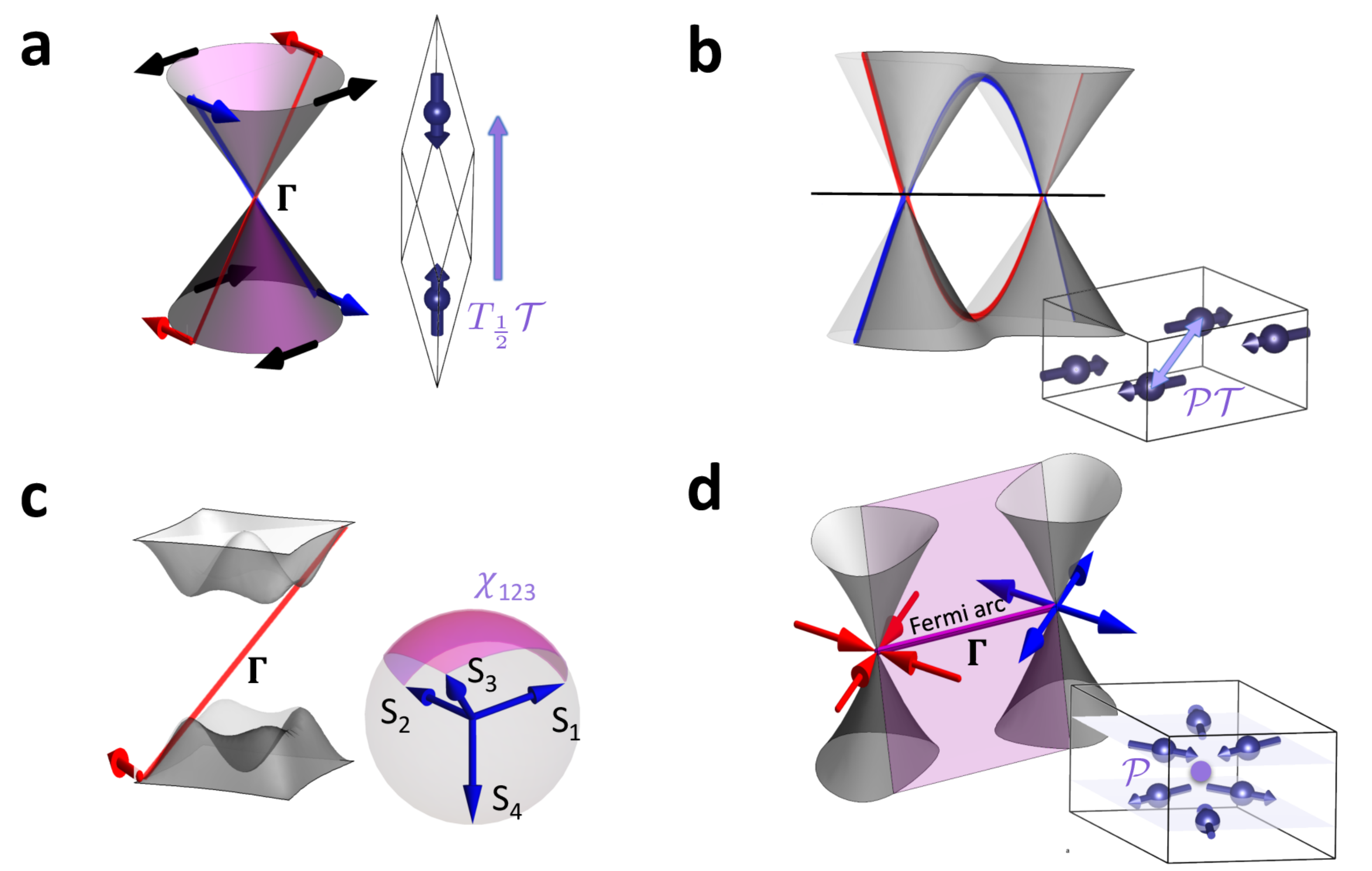}
\caption{\textbf{(BOX:) Overview of the materials palate of topological antiferromagnetic spintronics:} 
All panels show energy {\it vs.} crystal momentum band diagrams
and magnetic structures. 
 (a) Schematics of the spin-momentum locked surface-state Dirac quasiparticles in a TI. 
 In the inset, we show the magnetic Gd sublattice of GdPtBi, an AF TI candidate. 
 The magenta arrow marks the $T_{\frac{1}{2}}\mathcal{T}$ symmetry that protects a TI state in an AF. 
 (b) Schematics of the bulk Dirac quasiparticles of an AF Dirac semimetal (DSM) that must be located along a high symmetry line in the BZ, and the magnetic structure of the AF DSM candidate CuMnAs. 
 The $\mathcal{PT}$ symmetry connects the magnetic Mn sublattices in pairs,
 and together with additional crystalline symmetry protects the DSM state as we explain in Fig.\ref{Fig2}(b). 
(c) Massive Dirac quasiparticles and QAHE edge states (red dispersion). 
A quantized Hall conductivity can be produced by the spin-chirality, 
$\chi_{ijk}=\hat{S}_{i}\cdot \left(\hat{S}_{j} \times \hat{S}_{k}\right)$, of a non-coplanar spin texture. 
The quantized topological Hall effect (QTHE) in a non-coplanar insulating AF gives 
rise to quantized edge states as in the QAHE, as we explain in the text. 
For the sake of simplicity, the non-coplanar spins of the antiferromagnetic 
QTHE candidate K$_{\text{0.5}}$RhO$_{\text{2}}$ are translated to a common origin.
(d) Weyl points act as sources and drains of Berry curvature (blue and red arrows).  In a 
centrosymmetric WSM, the topological charges are distributed antisymmetrically around the $\Gamma$ point in the BZ.
The noncollinear magnetic sublattice of antiferromagnetic Mn$_{\text{3}}$Ge 
forms kagome planes and has inversion symmetry $\mathcal{P}$. 
The pseudorelativistic linear band crossings in the left panels are found in the edge/surface states, 
whereas those in the right panels are realized in the bulk. 
The TI and DSM states ( the two upper panels) cannot be realized in systems with
ferromagnetic order, but are possible in antiferromagnets.
In contrast, the QAHE/QTHE and WSM states  (the two lower panels) 
are expected in both FMs and AFs, and can be formally obtained by breaking symmetries of the phases depicted in the two upper panels. 
}
\label{Fig1}
\end{figure*}

As pointed out by Mong {\it et al.} \cite{Mong2010},
TI phases are possible in antiferromagnets even though time-reversal symmetry $\mathcal{T}$ 
is broken and are protected instead by $T_{\frac{1}{2}}\mathcal{T}$ where $T_{\frac{1}{2}}$ is a  half magnetic unit cell translation operation, as we illustrate in Fig.\ref{Fig1}(a). 
The proposed low-temperature AF candidate, GdPtBi, has 
not yet been confirmed as a TI by angle resolved photoemission 
spectroscopy (ARPES), presumably due imperfect
crystal momentum resolution of the measurement \cite{Liu2011f}. 
A path of research related to topological superconductivity has demonstrated signatures of the  
coexistence of a 2D TI, {\it i.e.} the quantum spin Hall effect (QSHE), and a 
superconducting state in hole-doped and electron-doped antiferromagnetic monolayers 
of FeSe \cite{Wang2016e}. FeSe is the metallic building block of the iron-based
high-T$_{\text{C}}$ superconductors. Surprisingly, the combined effect of substrate 
strain, spin-orbit coupling, and electronic correlations was shown to induce
band inversion and QSHE edge states \cite{Wang2016e}. 
These can in turn lead to the realization of Majorana zero modes, superconducting quasiparticle states 
with real wavefunctions that prevents decoherence, providing one route for the realization of 
quantum computing \cite{Beenakker2016,Tsai2016} with topological qubits.
The antiferromagnetic TI in combination with superconductivity can 
allow for an alternative realization of the Fu-Kane Majorana fermion proposal\cite{Fu2008}, possibly at higher temperatures \cite{Wang2016e}. 
Separately, QSHE states in an AF have also been 
predicted in honeycomb lattice systems \cite{Niu2017}.

\section{Topological semimetal antiferromagnets}

Topological semimetal states arise when conduction and valence bands touch at 
discrete points, lines, or planes in a bulk Brillouin zone at energies close to the Fermi level.
The low energy physics of topological semimetals is governed by 
effective Dirac or Weyl equations \cite{Burkov2016,Smejkal2017,Armitage2017}. 
3D Dirac and Weyl quasiparticles in non-magnetic bulk systems have attracted
attention because of reports of suppressed backscattering, 
measurements of exotic topological surface states,
and interest in unique topological responses such as dissipationless 
axial currents.\cite{Liang2014,Burkov2016,Jia2016}. 
These properties are thought to be responsible for experimental observations of chiral magnetotransport\cite{Hirschberger2016,Felser2016}, and 
strong magnetoresistance\cite{Ali2014,Soluyanov2015}, although the topological origin of these phenomena 
is not yet firmly established \cite{Liang2014}. 
For instance, the strong magnetoresistance in WTe$_{\text{2}}$ semimetals was originally 
explained on the basis on the carrier compensation in the tiny electron-hole pockets at the 
Fermi level \cite{Pletikosic2014}, and only later linked to the presence of Weyl 
fermions \cite{Soluyanov2015}.

\subsection{Topological metal-insulator transitions in 3D Dirac semimetal antiferromagnets}
In a system with time reversal $\mathcal{T}$ and spatial inversion $\mathcal{P}$ symmetries, the electronic bands are doubly degenerate resulting in a low energy Dirac Hamiltonian, $\mathcal{H}_{D}(\textbf{k})$.
In its simplest form\cite{Burkov2016,Smejkal2017,Armitage2017}: 
\begin{align}
\mathcal{H}_{D}(\textbf{k}) = 
\left(\begin{matrix} 
 \hbar v_{F}\textbf{k}\cdot \boldsymbol\sigma & m \\
m &  -\hbar v_{F}\textbf{k}\cdot \boldsymbol\sigma
\end{matrix}\right),
\label{MDirac}
 \end{align} 
where  $\boldsymbol\sigma$  is the vector of Pauli matrices, $v_{F}$ is the Fermi velocity, $\textbf{k}=\textbf{q}-\textbf{q}_{0}$ is the crystal momentum measured from the Dirac point at $\textbf{q}_{0}$, and $m$ is the mass (in units of energy). 
The corresponding energy dispersion reads 
$E_{\textbf{k}}=\pm \hbar v_{\text{F}}\sqrt{k_{x}^{2}+k_{y}^{2}+k_{z}^{2}+(\frac{m}{\hbar v_{F}}})^2$. 
The mass can vanish only when turned-off by a crystalline symmetry,
and in this case $\mathcal{H}_{D}(\textbf{k})$ describes the  
four-fold degenerate band touching \cite{Burkov2016,Smejkal2017} of a 
3D Dirac semimetal (DSM) illustrated schematically in Fig.\ref{Fig1}(b). 
In a 3D DSM, the topological invariants and 
nontrivial surface states can be linked to the crystalline symmetry protecting the degeneracy \cite{Yang2014a,Kargarian2016}.

\begin{figure*}[h!]
  \includegraphics[width=0.8\textwidth]{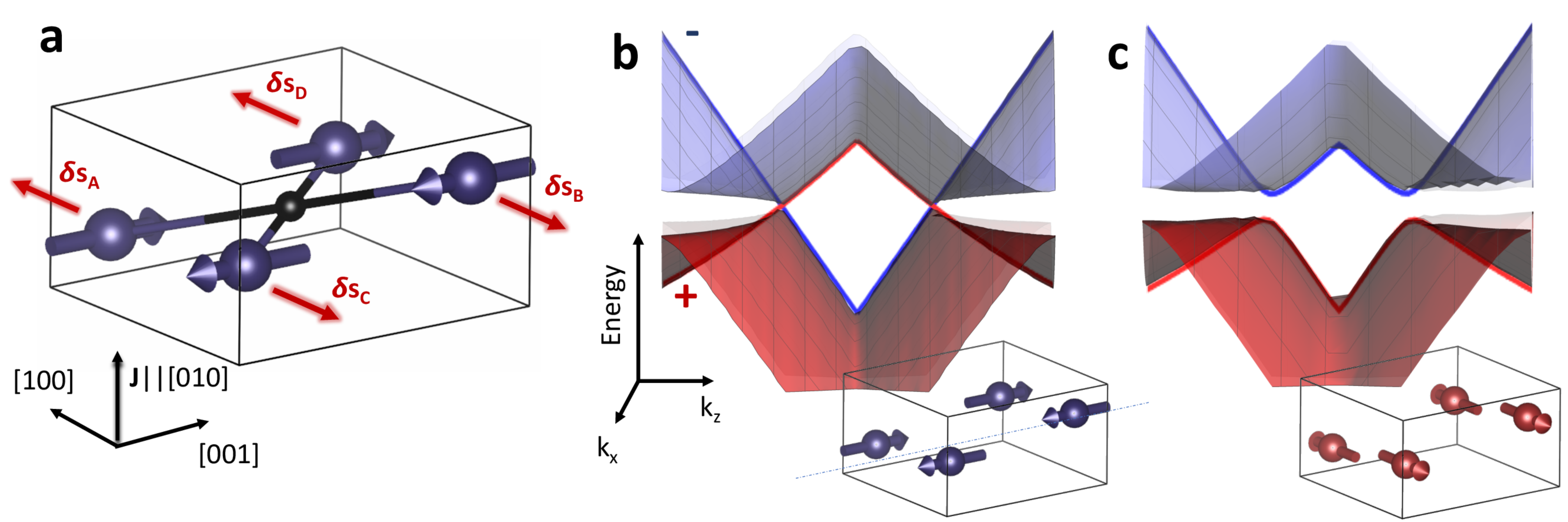}
\caption{\textbf{(BOX:) Topological metal-insulator transition in the antiferromagnetic DSM CuMnAs.} (a) The crystal and magnetic structure of the orthorhombic CuMnAs breaks time reversal and spatial inversion symmetry but preserves the combination $\mathcal{PT}$ which connects the $A-B$ and $C-D$ Mn 
sublattices (Cu and As atoms are omitted for brevity).
In the presence of an electrical current, owing to the $\mathcal{PT}$ symmetry, 
a non-equilibrium staggered spin-polarization, $\delta \textbf{s}$, is generated that
can reorient the antiferromagnetic moments. 
(b) For a N\'{e}el vector along the [001] axis, the crystal has a screw rotation 
symmetry $\mathcal{S}_{z}$ which transforms the atom $A$ into the atom 
$C$ by a $\pi$-rotation along the $z$-axis followed by a $(\frac{1}{2},0,\frac{1}{2})$-unit cell translation. 
$\mathcal{S}_{z}$ prevents hybridization of doubly degenerate bands that have 
opposite eigenvalues of $\mathcal{S}_{z}$ distinguished by $"+"$ and $"-"$ labels, 
and protects the DSM phase, as seen in the 
\textit{ab initio} band structure. 
(c) For the [100] orientation of the N\'{e}el vector, the $\mathcal{S}_{z}$ symmetry is broken 
and the Dirac fermions acquire a mass giving rise to a semiconducting phase. 
This N\'{e}el vector reorientation controlled transition is referred to as the topological metal-insulator transition and 
can lead to extremely large anisotropic magnetoresistance.
}
\label{Fig2}
\end{figure*}

The 3D DSM state is not possible in FMs because 
$\mathcal{T}$-symmetry breaking prevents double band degeneracy. 
On the other hand, a topological crystalline 3D Dirac semimetal was predicted in an AF, namely in the orthorhombic phase of
 CuMnAs \cite{Tang2016,Smejkal2016}. Here $\mathcal{P}$ and $\mathcal{T}$ 
symmetries are absent separately, but the combined effective $\mathcal{PT}$ symmetry ensures double band 
degeneracy over the whole Brillouin zone. The DSM is in this case protected by $\mathcal{PT}$ symmetry 
together with an additional crystalline non-symmorphic symmetry, as we explain in Fig.~\ref{Fig2}. 
The orthorhombic CuMnAs AF provides an attractive {\it hydrogen atom} for 
magnetic DSMs induced by the band inversion, 
since only a single pair of Dirac points appears near the Fermi level of the {\em ab initio} band structure. 
Electron filling enforced semimetals with a single Dirac cone are also 
a possibility as indicated theoretically in two dimensional model AFs \cite{Young2016}.

Novel effects have been predicted in topological DSM AFs that are based on the possibility of controlling topological states by controlling only N\'{e}el vector orientation,
not the presence or absence of antiferromagnetic order, and this can be accomplished 
using current-induced spin-orbit torques. 
The latter effect, discussed in detail by \v{Z}elezny {\em et al.} in this focused issue, has been experimentally 
demonstrated in  CuMnAs \cite{Wadley2016}. The coexistence of  Dirac fermions and spin-orbit torques in 
CuMnAs implies a new phase transition mechanism, referred to as the 
topological metal-insulator transition (MIT) \cite{Smejkal2016}. 
The origin of the effect is in Fermi surface topology that is sensitive to the changes in the magnetic 
symmetry upon reorienting the N\'{e}el vector, as explained in Fig.~\ref{Fig2}. 
The transport counterpart of the topological MIT is topological anisotropic magnetoresistance (AMR),
which in principle can reach extremely large values \cite{Smejkal2016}. The topological AMR can be understood as a 
limiting case of the crystalline AMR. The effect is different in origin and presumably more favorable for spintronics than the MIT observed in the pyrochlore iridate family which is driven by combined correlation and external field effects \cite{Tian2015}, or the extreme magnetoresistance observed in the AF topological metal 
candidate NdSb \cite{Wakeham2016}. 

An antiferromagnetic Dirac nodal line semimetal has also been proposed \cite{Smejkal2016}. 
Here the band touching lines are protected by off-centered mirror plane symmetries
and lead to distinct physical properties such as drum head surface states. 
Since the nodal-lines were observed several eV deep in the {\em ab initio} Fermi sea of 
tetragonal CuMnAs, the search is still on for more favorable candidate AF materials 
featuring nodal lines closer to the Fermi level.

\subsection{Weyl fermions in antiferromagnets}
When $\mathcal{P}$ or $\mathcal{T}$ symmetry, or both,
is broken and the double band degeneracy is lifted, the touching points of two non-degenerate bands can form a 3D Weyl semimetal (WSM), see Fig.\ref{Fig1}(d). 
Fermi states in a WSM are described by the Weyl Hamiltonian\cite{Burkov2016,Smejkal2017}: 
\begin{equation}
\mathcal{H}(\textbf{k})=\pm\hbar v_{\text{F}}\textbf{k}\cdot \boldsymbol\sigma.
\label{Weyl}
\end{equation}
Weyl points act as monopoles sources of Berry curvature flux and  
generate a topological charge defined by:
\begin{equation}
\mathcal{C}=\frac{1}{2\pi}\int_{\delta S}d^{2}k \; \textbf{n}\cdot\left\langle\partial_{\textbf{k}}u_{\textbf{k}}\vert\times \vert \partial_{\textbf{k}}u_{\textbf{k}} \right\rangle=\pm 1\, .
\label{Eq1}
\end{equation}
Here $\delta S$ is a small sphere surrounding the Weyl point with the surface normal vector $\textbf{n}$ and $\Omega_{\textbf{k}}=\langle\partial_{\textbf{k}}u_{\textbf{k}}\vert\times \vert \partial_{\textbf{k}}u_{\textbf{k}} \rangle $ is the momentum space Berry curvature, which can be viewed as a 
fictitious magnetic field \cite{Xiao2010b}. 
In the vicinity of the Weyl point the Berry curvature takes the 
monopole form, $\Omega_{\textbf{k}}=\pm \textbf{k}/ (2k^{3})$. 
Weyl points always come in pairs with opposite topological charges
and do not rely on any specific symmetry protection. 
The only way to remove them is to annihilate two Weyl points with opposite topological charges.
This is in contrast to the DSM case in which gaps 
can open also due to the hybridization among the degenerate band partners when 
symmetries are weakly broken, as we explained in the previous section. 
The 3D nature of the Weyl point is crucial here since the corresponding Weyl equation uses all three Pauli matrices.
Consequently, any small perturbation, that is in general expressed as a linear combination of Pauli matrices 
that form the basis of the $2\times2$ Hilbert space, just shifts but does not gap the Weyl point. (For example, for  a perturbation of a form $m\sigma_{z}$, the dispersion is renormalized as $E_{\textbf{k}}=\pm \hbar v_{\text{F}}\sqrt{k_{x}^{2}+k_{y}^{2}+(k_{z}+\frac{m}{\hbar v_{\text{F}}})^{2}}$.) 

\begin{figure*}[h!]
  \includegraphics*[width=0.8\textwidth]{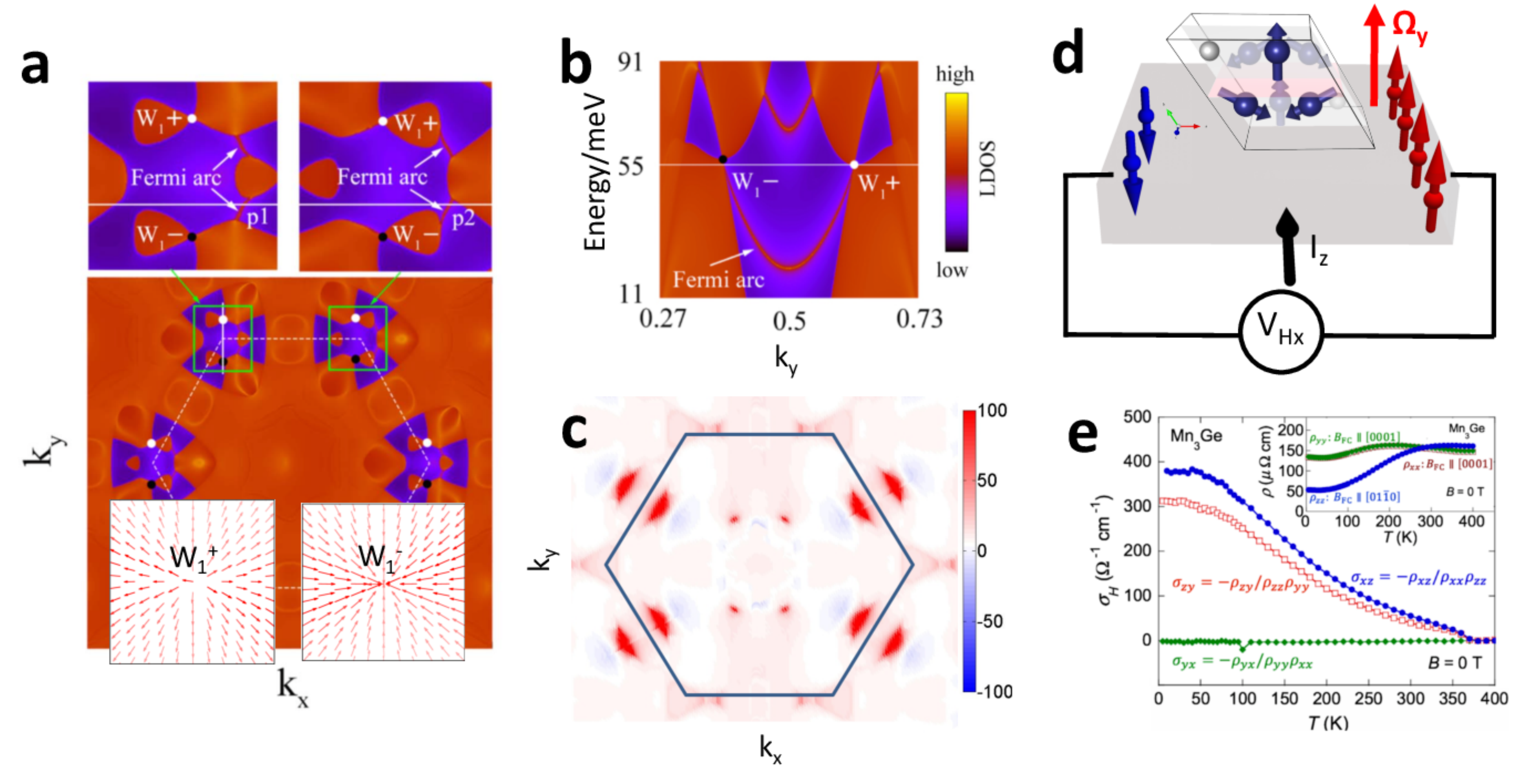}
\caption{\textbf{Weyl fermions and Hall effects in noncollinear AFs.}
(a) The density of states weighted by the surface contribution calculation reveals the Fermi surface with visible Fermi arcs. The Berry curvature vector field around the two Weyl points with opposite charges is plotted in two lower insets. 
(b) Weyl points in Mn$_{3}$Ge close to the Fermi level are connected by surface state Fermi arcs as calculated \textit{ab initio}.  
(c) The largest contribution to the intrinsic AHE originates from avoided crossing lines where the fictitious magnetic field, i.e. the Berry curvature, takes the largest value. 
(d) The concept of magnetic memory bits in a non-collinear AF Mn$_{3}$Ge. The measured large anomalous Hall effect at the room temperature originates in breaking the effective time reversal symmetry combining time reversal and mirror plane symmetries by the non-collinear magnetic order,  as shown in the inset on the magnetic structure of Mn$_{3}$Ge. The magnetic information can be stored in the orientation of the non-collinear AF order since a reorientation by, e.g., a weak magnetic field $\textbf{B}_{[010]}$ can flip the sign of the anomalous Hall voltage $V_{H}$ (and the corresponding conductivity $\sigma_{xz}$) for the applied current along the [001] direction. 
Temperature dependence of (e) the AHE conductivity in Mn$_{3}$Ge. 
The Figs.(a-b, and e) were adapted from \cite{Yang2016a,Kiyohara2015}.
}
\label{Fig3}
\end{figure*}

The initial prediction of the WSM by Wan et al\cite{Wan2011} was for pyrochlore AFs. Subsequently, a Weyl metal state was predicted in the chiral non-collinear centrosymmetric AF Mn$_{3}$Ge (see Fig.\ref{Fig1}(c)), 
a metallic system in which trivial Fermi surface pockets are also present\cite{Yang2016a}. 
Figs.\ref{Fig3}(a,b) illustrates the density of states weighted by the surface contribution which 
exhibits the typical Fermi arc features (see also Fig.\ref{Fig1}(d)), open surface state constant energy surfaces 
that connect the bulk projections of Weyl points, as found in \textit{ab initio} calculations. 
We note that Weyl semimetal states can be also be realized in the paramagnetic and 
AF phase of GdPtBi \cite{Hirschberger2016,Felser2016} by applying a magnetic field and, in contrast to 
DSMs, WSMs can in principle exist also in FMs \cite{Wang2016c}. 

The presence of Weyl points in the pyrochlore AF Eu$_{2}$Ir$_{2}$O$_{7}$ \cite{Sushkov2015} was inferred from optical experiments. However, a direct spectroscopic identification ({\it e.g.} by ARPES) of these Weyl AFs has yet to be made \cite{Yan2016a,Armitage2017}. 
The YbMnBi$_{2}$ AF was originally proposed to be WSM based on the ARPES identification of the bandcrossings consistent with the \textit{ab initio} bandstructure calculated for the canted AF moments \cite{Borisenko2015}, later supported by an optical study \cite{Chinotti2016}. However, recent magnetotransport\cite{Wang2016k} and optical conductivity\cite{Chaudhuri2017} measurements in the YbMnBi$_{2}$ AF were shown to be rather consistent with the Dirac qusiparticles \cite{Armitage2017}.
The Mn$_{3}$Ge AF was shown to host a large anomalous Hall effect (AHE) whose origin is discussed in the next 
section. 

\section{Hall effects in noncollinear topological antiferromagnets}

Until recently the AHE was viewed as a combined consequence of the time reversal symmetry breaking 
in a ferromagnet and spin-orbit coupling\cite{Sinova2015}. 
In the case of collinear AFs, either $T_{\frac{1}{2}}\mathcal{T}$ symmetry or $\mathcal{PT}$
symmetry forces the Hall conductivity to vanish.
Recent \textit{ab initio} calculations \cite{Chen2014,Kubler2014} inspired by earlier 
theoretical work\cite{Haldane1988,Shindou2001}\cite{Tomizawa2009,*Tomizawa2010}, have however shown that 
time-reversal symmetry breaking by AF order can yield a finite Hall response
in some non-collinear AFs, even those with zero net magnetization and even in the absence of 
spin-orbit coupling. The time reversal symmetry breaking is manifested by a nonzero Berry curvature, 
as we show in Fig.\ref{Fig3}(d)).  
The intrinsic contribution to the Hall conductivity depends only on the band structure of the perfect crystal 
and can be calculated from linear response theory \cite{Sinova2015}:
\begin{equation}
\sigma_{xz}=\frac{e^{2}}{\hbar}\sum_{n}\int_{\text{BZ}}\frac{d^{3}k}{(2\pi)^{3}} f(\textbf{k})\Omega^{n}_{y}(\textbf{k}),
\label{Eq2}
\end{equation}
where $\Omega_{y}^{n}(\textbf{k})$ is the $y$-component of the fictitious magnetic field, or the Berry curvature (cf. Eq.\eqref{Eq1}), and $n$ is the band index. 

\subsection{The Anomalous Hall effect in antiferromagnets}
In the simplest toy model of a WSM (with a two Weyl points in the vicinity of the Fermi level), the AHE conductivity can be calculated by integrating 
the quantized two-dimensional Hall conductivities of momentum-planes that are perpendicular 
to the line connecting Weyl points to obtain,
\begin{equation}
\sigma_{xy}=\frac{e^{2}}{h}\frac{\Delta k_{W}}{2\pi},
\end{equation}
where $\Delta k_{W}$ is the distance between Weyl points \cite{Burkov2011a,*Yang2011b}.
The AHE was recently observed in the hexagonal noncollinear AFs Mn$_{3}$Sn and Mn$_{3}$Ge \cite{Nakatsuji2015,Nayak2016,Kiyohara2015}, which have Weyl points close to the 
Fermi level.  However, \textit{ab initio} calculations of the intrinsic AHE in Mn$_{3}$Ge,
which predict a magnitude consistent with experiment (see Fig.\ref{Fig3}(e)),
reveal that the dominant contribution to the AHE comes instead from localized 
avoided crossings in the band structure \cite{Zhang2016d}.  This property 
is illustrated in Fig.\ref{Fig3}(c) by calculating crystal momentum projected contributions
of the Berry curvature/conductivity $\sim \int_{k_{z}}dk_{z}\Omega_{y}(k_{x},k_{y},k_{z})$.  
We also note that a large AHE was achieved in the collinear AF GdPtBi, 
by canting the staggered order \cite{Suzuki2016}. 

The discovery of a large AHE Mn$_{3}$Ge, which is a metal but has a relatively small density-of-states at the 
Fermi level, inspires a search for the quantized and dissipationless limits of the anomalous transport in topological semiconducting/insulating AFs.  A quantum anomalous Hall effect (QAHE) with quantized charged edge state channels
was proposed for antiferromagnetic insulators \cite{Zhou2016a,Dong2016}.  
In a two-band 2D Chern insulator with 
Hamiltonian $\mathcal{H}_{\textbf{k}}=\textbf{d}(\textbf{k}) \cdot \boldsymbol\sigma$,
 the 2D variant of  Eq.\eqref{Eq2} reads
\begin{equation}
\sigma_{xy}=\frac{e^{2}}{h}\frac{1}{4\pi}\sum_{n}\int_{\text{BZ}}dk_{x}dk_{y} \hat{\textbf{d}}\cdot \left( \partial_{k_{x}}\hat{\textbf{d}} \times \partial_{k_{y}}\hat{\textbf{d}} \right),
\label{CHI}
\end{equation}
where $\hat{\textbf{d}}(\textbf{k})=\textbf{d}(\textbf{k})/\vert \textbf{d}(\textbf{k}) \vert$.
The integral is quantized and relates directly to the Chern number (c.f. Eq.\eqref{Eq1}).  

While the AHE arises from the Berry curvature in the momentum space, other important spintronic 
phenomena can be associated with Berry curvatures in different parameter spaces.  
For instance, the spin-orbit torkance tensor is defined by the linear response
relation $\textbf{T}=\tau\textbf{E}$, where $\textbf{T}=\frac{d\textbf{m}}{dt}$ is the SOT exerted on the magnetization $\textbf{m}$ in a magnet subject to an applied electric field $\textbf{E}$. The intrinsic part of the SOT can be rewritten in terms of a mixed Berry curvature, $\Omega_{ij}^{\hat{\textbf{m}}\textbf{k}}=\hat{\textbf{e}}_{i}\cdot 2\text{Im}\sum_{n}\left\langle \partial_{\hat{\textbf{m}}}u_{\textbf{k}n}\vert \partial_{k_{j}}u_{\textbf{k}n}\right\rangle$, where $\hat{\textbf{e}}_i$ denotes the $i$th Cartesian unit vector and $\hat{\textbf{m}}$ is a unit vector in the direction of magnetization \cite{Hanke2017a}. A large SOT in a topologically nontrivial insulating FM has been
associated with the existence of the monopoles of the mixed Berry curvature. 
These are termed mixed Weyl points as they correspond formally to a Weyl Hamiltonian $\mathcal{H}(\textbf{k},\hat{\textbf{m}})=\hbar v_{F}\left( k_{x}\sigma_{x}+k_{y}\sigma_{y} \right)+v_{\theta}\theta \sigma_{z}$ in the mixed momentum-magnetization space\cite{Hanke2017a}. 
(Here $\theta$ is the azimuthal angle of the magnetization.) 
The recent discovery of the SOT and the prediction of the DSM in antiferromagnetic 
CuMnAs motivates a search for analogous dissipationless (pronounced) SOTs in insulating AFs.  
Finally, we note that spintronics based on insulating AFs may 
utilize the concept of Weyl magnons (a boson analog of Weyl fermions), 
that has been proposed in pyrochlore AFs\cite{Li2016a}.

\subsection{Topological Hall effect in antiferromagnets} 
Real-space order parameter textures can be induced in AFs, and their presence can be detected by the so-called topological Hall effect. 
In this phenomenon, the role of the spin-orbit coupling is substituted by the chirality of the spin 
texture (see inset in Fig.\ref{Fig1}(c)), $\chi_{ijk}=\hat{S}_{i}\cdot \left(\hat{S}_{j} \times \hat{S}_{k}\right)$. 
(Note that the spin-chirality vanishes in coplanar AFs like Mn$_{3}$Ge.)  
The effect of the corresponding fictitious magnetic field, $\hat{\textbf{m}}\cdot \left(\partial_{x}\hat{\textbf{m}} \times \partial_{y}\hat{\textbf{m}}\right)$,  on the Bloch electrons generates a Hall response. The topological Hall effect can be experimentally distinguished from the AHE by, e.g., analyzing the disorder dependence \cite{Kanazawa2011}. However, the distinction at surfaces might be difficult as was pointed out in studies of monolayer Fe deposited on an Ir(001) surface \cite{Hoffmann2015}. 

The Hall effect associated with the spin-chirality  was initially reported in antiferromagnetic pyrochlore iridates \cite{Machida2010} and later in MnSi chiral antiferromagnetic alloys \cite{Surgers2014,Surgers2016}. We note that the term "topological" used to label the effect does not imply in this case a correspondence to a topological invariant. This is in contrast to skyrmions
 discussed in more detail in the following section. A skyrmion spin texture carries an integer topological charge which is accompanied with a  topological Hall effect. In this case, the term topological refers to the association of the Hall response to a topological invariant. Another example of such a correspondence is the quantum topological Hall effect which was proposed for the non-coplanar AF  K$_{0.5}$RhO$_{2}$ \cite{Zhou2016}, as we show schematically in Fig.\ref{Fig1}(d). Unlike skyrmions, here the topological charge occurs in the momentum space, i.e., is obtained from Eq.\eqref{Eq2} giving an integer value. 

Finally, topological systems were predicted also as promising generators of the spin Hall effect (SHE). The WSM TaAs was predicted to host a large spin Hall angle arising mainly from the nodal line anticrossing features \cite{Sun2016a}. Similarly, a topological SHE was predicted for skyrmions\cite{Yin2015} where the spin Hall response occurs even in the absence of the spin-orbit coupling, in analogy with the above topological (charge) Hall effect. Recently, it was theoretically proposed that the SHE can occur also due to the breaking of the spin rotational symmetry in non-collinear AFs without the need for either the spin-orbit coupling or the spin-chirality \cite{Zhang2017b}.

In noncollinear AFs, other "topological" phenomena might arise, such as the topological orbital magnetism in  Fe/Ir(001), Mn/Cu(111) or in the $\gamma$-phase of FeMn alloy \cite{Hoffmann2015,Hanke2016,Hanke2017b}. The effect is rooted in the spin-chirality of non-coplanar spin textures, exists also without the spin-orbit coupling, and does not have to be associated with an integer topological invariant \cite{Hanke2017b}. Novel functionalities can be envisaged by controlling the topological orbital magnetism via spin-torque manipulation of the spin textures or via the spin-orbit coupling \cite{Hanke2017b}.

\section{Antiferromagnetic skyrmions}

\begin{figure*}[t]
  \includegraphics*[width=0.6\textwidth]{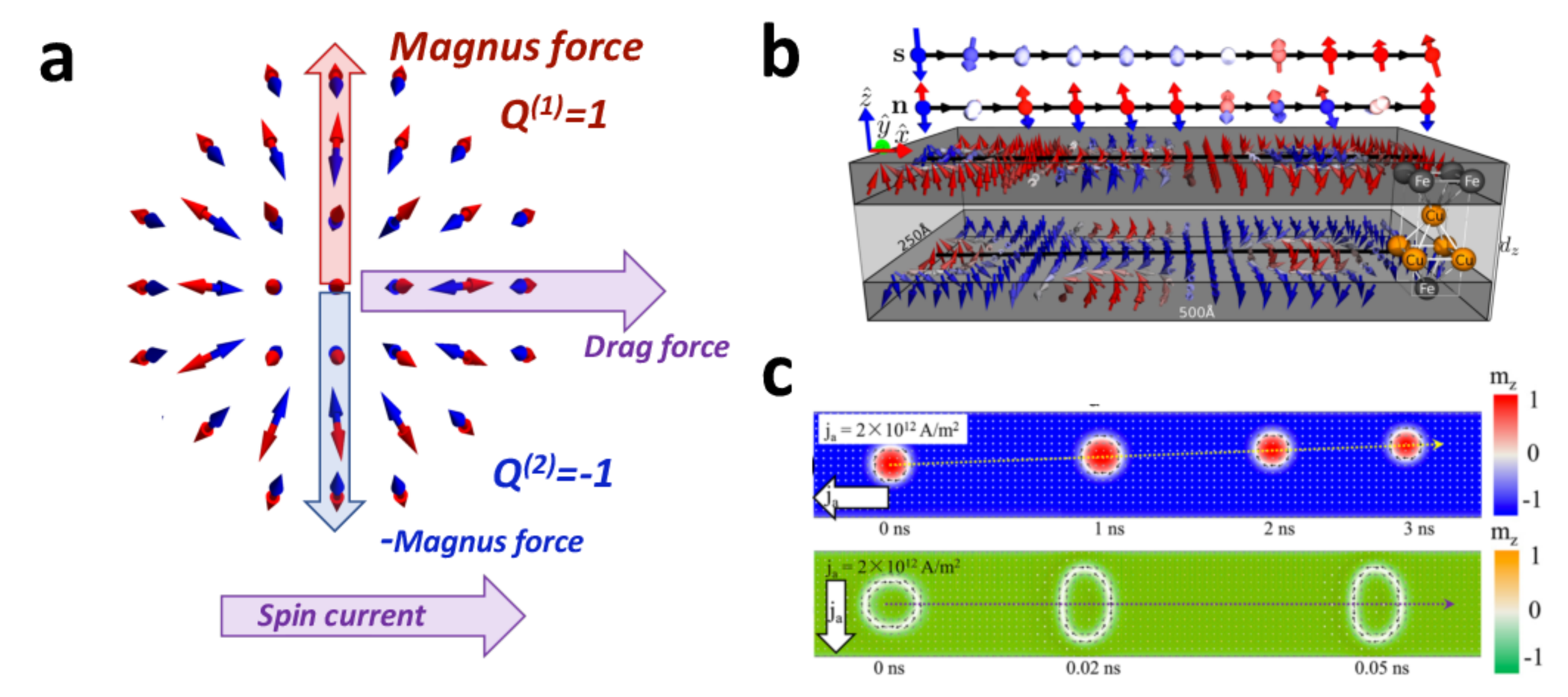}
\caption{\textbf{Antiferromagnetic skyrmions.} (a) An antiferromagnetic skyrmion can be viewed as consisting of two antiferromagnetically coupled ferromagnetic skyrmions. For the sake of clarity, we draw the two opposite magnetic moments in the antiferromagnetic unit cell as coinciding and their moments as perfectly compensated. 
Note that the structure of the skyrmion is analogous to the momentum-space Berry curvature shown in the inset of Fig.\ref{Fig3}(a). (b) A synthetic antiferromagnetic skyrmion in a Fe-Cu-Fe trilayer. (c) Micromagnetic simulation of ferromagnetic (upper panel) and antiferromagnetic (lower panel) skyrmion motion driven by a SOT. 
The ferromagnetic skyrmion is deflected by the Magnus force, while the antiferromagnetic skyrmion can move in a straight line due to mutual compensation between Magnus forces from the two magnetic sublattices, as schematically shown in panel (a). Panels (b)-(c) are adapted from Refs. \onlinecite{Buhl2017,Jin2016}.  
}
\label{Fig4}
\end{figure*}

As noted above, position-dependent magnetization textures can also be topologically non-trivial.
Skyrmions are noncollinear magnetization textures in which 
the spin quantization axis changes continuously over length scales that 
vary from a few nm to a few $\mu$m.  For two-dimensional systems, the winding number,  
\begin{equation}
Q^{(j)}=\frac{1}{4\pi}\int dxdy \, \hat{\textbf{m}}^{(j)}\cdot \left(\partial_{x}\hat{\textbf{m}}^{(j)} \times \partial_{y}\hat{\textbf{m}}^{(j)}\right),
\label{Eq3}
\end{equation}
of a magnetization texture measures the number of times the sphere of magnetization 
directions is covered upon integrating over 
space and must take on integer values\cite{Fert2013,Finocchio2016}.
Here $\hat{\textbf{m}}=\textbf{m}(x,y,z)$ is the normalised magnetization field in the real space and  $\hat{\textbf{m}}\cdot \left(\partial_{x}\hat{\textbf{m}} \times \partial_{y}\hat{\textbf{m}}\right)$ is the fictitious emergent magnetic field.  The antiferromagnetic skyrmion can be visualized as two interpenetrating ferromagnetic skyrmions, where the index $(j)=(1,2)$ labels the two antiferromagnetic sublattices,  as shown in Fig.\ref{Fig4}(a). Microscopically, the skyrmionic magnetization modulation is caused by the Dzyaloshinskii-Moriya interaction (DMI) of non-centrosymmetric crystals. Whereas  ferromagnetic skyrmions are often generated by interfacial DMIs, antiferromagnetic skyrmions are 
expected to be more abundant in crystals with bulk DMI \cite{Liu2016a}. 

By comparing to Eq.\eqref{Eq1}, and \eqref{CHI} we see that the winding number $Q$ topologically protects skyrmion textures 
in real space, just as Weyl points and the QAHE state are protected in momentum space. 
The observed energy barrier for skyrmion annihilation in discrete magnetic skyrmions
is of the order of $\sim$0.1~eV \cite{Rohart2016}. 
Because this barrier is finite, skyrmion stability in experimental systems relies in part on 
other physical limitations, ({\it e.g.} a combined effect of spin rotation and skyrmion diameter shrinking)  rather than from
topological protection itself \cite{Rohart2016}.

Skyrmions in non-centrosymmetric AFs were considered already some time ago \cite{Bogdanov2002} in various contexts, including in connection with high-temperature superconductivity \cite{Morinari2010}. 
However, spintronics aspects of antiferromagnetic skyrmions, namely their manipulation by an electrical current, were discussed only recently \cite{Zhang2015b,Barker2016,Jin2016,Velkov2016}.
Micromagnetic simulations show that antiferromagnetic skyrmions move faster than ferromagnetic skyrmions,
can be driven with lower current densities and, most importantly, move in straight lines, as illustrated in Fig.~\ref{Fig4}(c)\cite{Zhang2015b,Barker2016,Jin2016}. 
This important difference arises because the Magnus force which
deflects ferromagnetic skyrmions has opposite sign for 
the two magnetic sublattices of an antiferromagnetic skyrmion, 
owing to the opposite topological numbers illustrated in Fig.\ref{Fig4}(a). 
AF skyrmions were recently studied in detail also in synthetic AFs (e.g. in a Fe$\vert$Cu$\vert$Fe trilayer) 
in which skyrmions in the two ferromagnetic layers are 
coupled antiferromagnetically\cite{Zhang2016h,Buhl2017}, as shown in Fig.~\ref{Fig4}(b). 
The topological SHE was suggested as a probe to monitor the AF skyrmions,
as well as to generate a spin current \cite{Buhl2017}. 

\section{perspectives}

\begin{figure*}[t]
  \includegraphics*[width=0.6\textwidth]{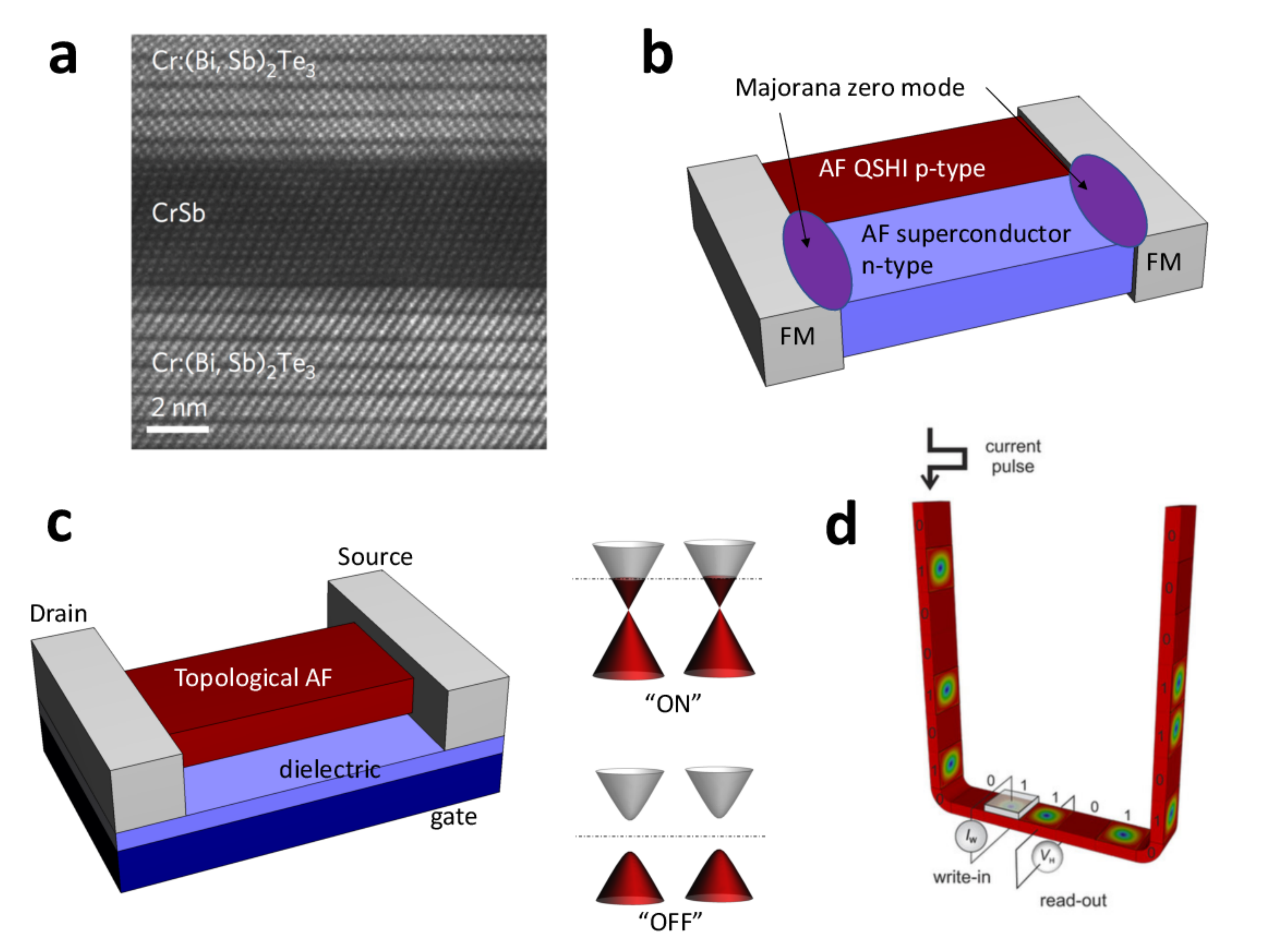}
\caption{\textbf{Topological spintronics based on antiferromagnets.} (a) Example of a possible 
building block for topological AF spintronics in heterostructures: an epitaxially matched interface between 
AF CrSb and a Cr-doped MTI. (b) A Majorana qubits in an
AF FeSe monolayer. (c) A concept of a topological transistor explained in the text. (d) The skyrmion racetrack memory concept.  Magnetic information is stored in a 
topological charge Q  instead of the magnetization of the magnetic domain. "1" encodes the skyrmion with Q=1, while "0" corresponds to a uniform domain with Q=0.  Spin transfer torque shifts the register. 
Bits read by the topological Hall effect and written by nucleating or deleting the skyrmion. 
 Panels (a), and (d) are adapted from Refs. \onlinecite{He2016,Zhang2015k}.  
}
\label{Fig5}
\end{figure*}

The past year has seen important progress in 
coupling topological phases of matter with AF order.
The fortunate lattice constant match between 
the Cr-doped TI  (Bi,Sb)$_{\text{2}}$Te$_{\text{3}}$, and the 
high temperature AF CrSb (see Fig.\ref{Fig5}(a)) has been exploited to grow
epitaxial interfaces between these materials.  The resulting heterostructures exhibit
strengthened order in the MTI \cite{He2016}, enhancing topological effects in its electronic properties.
CrSb/(Bi,Sb)$_{\text{2}}$Te$_{\text{3}}$ /CrSb trilayers exhibit cusps in 
the magnetoresistance, that presumably correspond to 
a topological phase transition of Dirac quasiparticles at the interfaces \cite{He2016a}. 
Separately, in studies of Bi$_{\text{2}}$Se$_{\text{3}}$/CoTb 
heterostructures \cite{Han2017}, it was also shown that a SOT from a TI
can manipulate the AF-coupled sublattices in an adjacent ferrimagnet.   
Further progress in which perfectly compensated antiferromagnetic materials are employed can enable the advantages of AF spintronics, discussed throughout this 
issue, to be realized more fully.  AF FeSe monolayers will likely be tested for the realization of
Majorana-based topological quantum qubits. 
Making a p-n junction in a  single layer of FeSe by gating can 
generate two regions - one superconducting and one with a QSHE \cite{Tsai2016}. 
Coupling this system to two FM electrodes from both sides would lead to localization of 
Majorana modes at the interface as illustrated in Fig.\ref{Fig5}(b). 
This two level state can function as a quantum bit which can nonlocally encode information and is robust against decoherence due to the real wavefunction of the Majorana modes \cite{Beenakker2016}. 

In this brief review, we have focused on direct synergies between antiferromagnetic 
and topological properties in crystal momentum and real spaces. 
Novel magnetic topological phases of matter were predicted 
only very recently and in many cases remain experimentally elusive, 
{\it e.g.} AF TIs, DSMs, and WSMs, QAHE states, and skyrmions. 
When realized, these topological antiferromagnet states
can lead not only to more stable topological 
nanospintronics devices, but also to unprecedented functionalities relying on the unique AF symmetries and
the possibility of tuning them by external means by coupling to the AF order.
Fast topological memories, in which states are written by the topological SOT in an AF TI, or an AF DSM \cite{Smejkal2016,Ghosh2017,Hanke2017a}, and read out 
via the large magnetoresistance effects associated 
with band gap tuning\cite{Smejkal2016,Kandala2015,Carbone2016}, are among anticipated applications.
Another possibility is opened by exploiting topological phase transitions, 
as we have explained for the CuMnAs Dirac AF in the Fig.\ref{Fig2}. 
Here one can foresee the possibility of a topological transistor operating at high frequencies and low current densities (see Fig.\ref{Fig5}(c)) \cite{Xue2011}.
In contrast to the related proposal in crystalline topological insulators \cite{Liu2013d}, one can achieve 
highly mobile bulk Dirac quasiparticle current (present in the "ON" state) controllable by gating or by ultrafast SOT. 
We note that many of the novel effects we have 
discussed follow directly from antiferromagnetic symmetries and cannot be realized in FMs, 
for instance (i) magnetism combined with the QSHE, and superconductivity, and (ii) magnetism combined 
with Dirac semimetal phase.
The conditions for a good Dirac quasiparticle
in AF spintronics were discussed recently \cite{Smejkal2017}.

The sign and magnitude of the AHE in Mn$_{\text{3}}$Ge depends on the noncollinear spin texture orientation. 
This together with the demonstration of the possibility of manipulating the noncollinear spin texture by a spin-torque\cite{Fujita2016} can allow for memory devices in noncollinear AFs,
with the electrical readout via the AHE, as illustrated in Fig.\ref{Fig3}(d). Moreover, optical counterparts of the dc AHE should be present in non-collinear AFs\cite{Feng2015}, opening the prospect for optical detection of 
topological effects and of antiferromagnetic opto-spintronic devices. 

The nontrivial topologies of magnetization texture might also find applications. 
The skyrmion might represent the smallest micromagnetic configuration for storing information before going into truly quantum mechanical single spin qubits \cite{Soumyanarayanan2016}. For instance, in the skyrmionic racetrack memory, as shown in Fig.~\ref{Fig5}(d), the magnetic information is stored in skyrmions instead of magnetic domains separated by domains walls \cite{Fert2013}. Skyrmions can be driven by the SOT  at low current densities and have advantages over domain walls especially in the curved parts of the race track thanks to their stability.       

After almost a half century of research we have discovered many manifestations of 
topology playing a role in materials physics, from spin liquids, to 
Quantum Hall effects, to the theory of dislocations \cite{thouless1998topological} and beyond.  
In this article, we have highlighted a newly emerging field, topological antiferromagnetic
spintronics.  Beyond providing an interesting new context in which to identify and understand the 
physical consequences of topological properties in momentum-space bands or 
real-space textures, topological antiferromagnetic spintronics 
suggests the tantalizing possibility of converting important fundamental advances to 
truly valuable new applications of quantum materials.

\textbf{Acknowledgement}
L\v{S} acknowledges support from the Grant Agency of the Charles University,  no. 280815. Access to computing and storage facilities owned by parties and projects contributing to the National Grid Infrastructure MetaCentrum provided under the program “Projects of Large Research, Development, and Innovations Infrastructures” (CESNET LM2015042) is greatly appreciated. YM acknowledges funding from the German Research Foundation (Deutsche Forschungsgemeinschaft, Grant No. MO 1731/5-1). B.Y. acknowledges support of the Ruth and Herman Albert Scholars Program for New Scientists in Weizmann Institute of Science, Israel.
AHM was supported by SHINES, an Energy Frontier Research Center funded by the U.S. Department of Energy, Office of Science, Basic Energy Sciences under Award \#SC0012670, 
and by Welch Foundation grant TBF1473.


\begin{thebibliography}{102}%
\makeatletter
\providecommand \@ifxundefined [1]{%
 \@ifx{#1\undefined}
}%
\providecommand \@ifnum [1]{%
 \ifnum #1\expandafter \@firstoftwo
 \else \expandafter \@secondoftwo
 \fi
}%
\providecommand \@ifx [1]{%
 \ifx #1\expandafter \@firstoftwo
 \else \expandafter \@secondoftwo
 \fi
}%
\providecommand \natexlab [1]{#1}%
\providecommand \enquote  [1]{``#1''}%
\providecommand \bibnamefont  [1]{#1}%
\providecommand \bibfnamefont [1]{#1}%
\providecommand \citenamefont [1]{#1}%
\providecommand \href@noop [0]{\@secondoftwo}%
\providecommand \href [0]{\begingroup \@sanitize@url \@href}%
\providecommand \@href[1]{\@@startlink{#1}\@@href}%
\providecommand \@@href[1]{\endgroup#1\@@endlink}%
\providecommand \@sanitize@url [0]{\catcode `\\12\catcode `\$12\catcode
  `\&12\catcode `\#12\catcode `\^12\catcode `\_12\catcode `\%12\relax}%
\providecommand \@@startlink[1]{}%
\providecommand \@@endlink[0]{}%
\providecommand \url  [0]{\begingroup\@sanitize@url \@url }%
\providecommand \@url [1]{\endgroup\@href {#1}{\urlprefix }}%
\providecommand \urlprefix  [0]{URL }%
\providecommand \Eprint [0]{\href }%
\providecommand \doibase [0]{http://dx.doi.org/}%
\providecommand \selectlanguage [0]{\@gobble}%
\providecommand \bibinfo  [0]{\@secondoftwo}%
\providecommand \bibfield  [0]{\@secondoftwo}%
\providecommand \translation [1]{[#1]}%
\providecommand \BibitemOpen [0]{}%
\providecommand \bibitemStop [0]{}%
\providecommand \bibitemNoStop [0]{.\EOS\space}%
\providecommand \EOS [0]{\spacefactor3000\relax}%
\providecommand \BibitemShut  [1]{\csname bibitem#1\endcsname}%
\let\auto@bib@innerbib\@empty
\bibitem [{\citenamefont {Sarma}, \citenamefont {Freedman},\ and\ \citenamefont
  {Nayak}(2015)}]{Sarma2015}%
  \BibitemOpen
  \bibfield  {author} {\bibinfo {author} {\bibfnamefont {S.~D.}\ \bibnamefont
  {Sarma}}, \bibinfo {author} {\bibfnamefont {M.}~\bibnamefont {Freedman}}, \
  and\ \bibinfo {author} {\bibfnamefont {C.}~\bibnamefont {Nayak}},\ }\bibfield
   {title} {\enquote {\bibinfo {title} {{Majorana Zero Modes and Topological
  Quantum Computation}},}\ }\href {\doibase 10.1038/npjqi.2015.1} {\bibfield
  {journal} {\bibinfo  {journal} {Nat. Publ. Gr.}\ } (\bibinfo {year} {2015}),\
  10.1038/npjqi.2015.1},\ \Eprint {http://arxiv.org/abs/1501.02813}
  {arXiv:1501.02813} \BibitemShut {NoStop}%
\bibitem [{\citenamefont {Beenakker}\ and\ \citenamefont
  {Kouwenhoven}(2016)}]{Beenakker2016}%
  \BibitemOpen
  \bibfield  {author} {\bibinfo {author} {\bibfnamefont {C.~W.~J.}\
  \bibnamefont {Beenakker}}\ and\ \bibinfo {author} {\bibfnamefont
  {L.}~\bibnamefont {Kouwenhoven}},\ }\bibfield  {title} {\enquote {\bibinfo
  {title} {{A road to reality with topological superconductors}},}\ }\href
  {\doibase 10.1038/nphys3778} {\bibfield  {journal} {\bibinfo  {journal} {Nat.
  Phys.}\ }\textbf {\bibinfo {volume} {12}},\ \bibinfo {pages} {618--621}
  (\bibinfo {year} {2016})},\ \Eprint {http://arxiv.org/abs/1606.09439v1}
  {arXiv:1606.09439v1} \BibitemShut {NoStop}%
\bibitem [{\citenamefont {Hasan}\ and\ \citenamefont {Kane}(2010)}]{Hasan2010}%
  \BibitemOpen
  \bibfield  {author} {\bibinfo {author} {\bibfnamefont {M.~Z.}\ \bibnamefont
  {Hasan}}\ and\ \bibinfo {author} {\bibfnamefont {C.}~\bibnamefont {Kane}},\
  }\bibfield  {title} {\enquote {\bibinfo {title} {{Colloquium: Topological
  insulators}},}\ }\href {\doibase 10.1103/RevModPhys.82.3045} {\bibfield
  {journal} {\bibinfo  {journal} {Rev. Mod. Phys.}\ }\textbf {\bibinfo {volume}
  {82}},\ \bibinfo {pages} {3045--3067} (\bibinfo {year} {2010})}\BibitemShut
  {NoStop}%
\bibitem [{\citenamefont {Fan}\ and\ \citenamefont {Wang}(2016)}]{Fan2016b}%
  \BibitemOpen
  \bibfield  {author} {\bibinfo {author} {\bibfnamefont {Y.}~\bibnamefont
  {Fan}}\ and\ \bibinfo {author} {\bibfnamefont {K.~L.}\ \bibnamefont {Wang}},\
  }\bibfield  {title} {\enquote {\bibinfo {title} {{Spintronics Based on
  Topological Insulators}},}\ }\href {\doibase 10.1142/S2010324716400014}
  {\bibfield  {journal} {\bibinfo  {journal} {SPIN}\ }\textbf {\bibinfo
  {volume} {06}},\ \bibinfo {pages} {1640001} (\bibinfo {year}
  {2016})}\BibitemShut {NoStop}%
\bibitem [{\citenamefont {Wang}\ \emph
  {et~al.}(2016{\natexlab{a}})\citenamefont {Wang}, \citenamefont {Kally},
  \citenamefont {Lee}, \citenamefont {Liu}, \citenamefont {Chang},
  \citenamefont {Hickey}, \citenamefont {Mkhoyan}, \citenamefont {Wu},
  \citenamefont {Richardella},\ and\ \citenamefont {Samarth}}]{Wang2016d}%
  \BibitemOpen
  \bibfield  {author} {\bibinfo {author} {\bibfnamefont {H.}~\bibnamefont
  {Wang}}, \bibinfo {author} {\bibfnamefont {J.}~\bibnamefont {Kally}},
  \bibinfo {author} {\bibfnamefont {J.~S.}\ \bibnamefont {Lee}}, \bibinfo
  {author} {\bibfnamefont {T.}~\bibnamefont {Liu}}, \bibinfo {author}
  {\bibfnamefont {H.}~\bibnamefont {Chang}}, \bibinfo {author} {\bibfnamefont
  {D.~R.}\ \bibnamefont {Hickey}}, \bibinfo {author} {\bibfnamefont {K.~A.}\
  \bibnamefont {Mkhoyan}}, \bibinfo {author} {\bibfnamefont {M.}~\bibnamefont
  {Wu}}, \bibinfo {author} {\bibfnamefont {A.}~\bibnamefont {Richardella}}, \
  and\ \bibinfo {author} {\bibfnamefont {N.}~\bibnamefont {Samarth}},\
  }\bibfield  {title} {\enquote {\bibinfo {title} {{Surface-State-Dominated
  Spin-Charge Current Conversion in
  Topological-Insulator–Ferromagnetic-Insulator Heterostructures}},}\ }\href
  {\doibase 10.1103/PhysRevLett.117.076601} {\bibfield  {journal} {\bibinfo
  {journal} {Phys. Rev. Lett.}\ }\textbf {\bibinfo {volume} {117}},\ \bibinfo
  {pages} {076601} (\bibinfo {year} {2016}{\natexlab{a}})}\BibitemShut
  {NoStop}%
\bibitem [{\citenamefont {Soumyanarayanan}\ \emph {et~al.}(2016)\citenamefont
  {Soumyanarayanan}, \citenamefont {Reyren}, \citenamefont {Fert},\ and\
  \citenamefont {Panagopoulos}}]{Soumyanarayanan2016}%
  \BibitemOpen
  \bibfield  {author} {\bibinfo {author} {\bibfnamefont {A.}~\bibnamefont
  {Soumyanarayanan}}, \bibinfo {author} {\bibfnamefont {N.}~\bibnamefont
  {Reyren}}, \bibinfo {author} {\bibfnamefont {A.}~\bibnamefont {Fert}}, \ and\
  \bibinfo {author} {\bibfnamefont {C.}~\bibnamefont {Panagopoulos}},\
  }\bibfield  {title} {\enquote {\bibinfo {title} {{Spin-Orbit Coupling Induced
  Emergent Phenomena at Surfaces and Interfaces}},}\ }\href {\doibase
  10.1038/nature19820} {\bibfield  {journal} {\bibinfo  {journal} {Nature}\
  }\textbf {\bibinfo {volume} {539}},\ \bibinfo {pages} {509--517} (\bibinfo
  {year} {2016})},\ \Eprint {http://arxiv.org/abs/1611.09521}
  {arXiv:1611.09521} \BibitemShut {NoStop}%
\bibitem [{\citenamefont {Pesin}\ and\ \citenamefont
  {MacDonald}(2012)}]{Pesin2012b}%
  \BibitemOpen
  \bibfield  {author} {\bibinfo {author} {\bibfnamefont {D.~A.}\ \bibnamefont
  {Pesin}}\ and\ \bibinfo {author} {\bibfnamefont {A.~H.}\ \bibnamefont
  {MacDonald}},\ }\bibfield  {title} {\enquote {\bibinfo {title} {{Spintronics
  and pseudospintronics in graphene and topological insulators.}}}\ }\href
  {\doibase 10.1038/nmat3305} {\bibfield  {journal} {\bibinfo  {journal} {Nat.
  Mater.}\ }\textbf {\bibinfo {volume} {11}},\ \bibinfo {pages} {409--416}
  (\bibinfo {year} {2012})}\BibitemShut {NoStop}%
\bibitem [{\citenamefont {Liang}\ \emph {et~al.}(2014)\citenamefont {Liang},
  \citenamefont {Gibson}, \citenamefont {Ali}, \citenamefont {Liu},
  \citenamefont {Cava},\ and\ \citenamefont {Ong}}]{Liang2014}%
  \BibitemOpen
  \bibfield  {author} {\bibinfo {author} {\bibfnamefont {T.}~\bibnamefont
  {Liang}}, \bibinfo {author} {\bibfnamefont {Q.}~\bibnamefont {Gibson}},
  \bibinfo {author} {\bibfnamefont {M.~N.}\ \bibnamefont {Ali}}, \bibinfo
  {author} {\bibfnamefont {M.}~\bibnamefont {Liu}}, \bibinfo {author}
  {\bibfnamefont {R.~J.}\ \bibnamefont {Cava}}, \ and\ \bibinfo {author}
  {\bibfnamefont {N.~P.}\ \bibnamefont {Ong}},\ }\bibfield  {title} {\enquote
  {\bibinfo {title} {{Ultrahigh mobility and giant magnetoresistance in the
  Dirac semimetal Cd3As2}},}\ }\href {\doibase 10.1038/nmat4143} {\bibfield
  {journal} {\bibinfo  {journal} {Nat. Mater.}\ }\textbf {\bibinfo {volume}
  {14}},\ \bibinfo {pages} {280--284} (\bibinfo {year} {2014})}\BibitemShut
  {NoStop}%
\bibitem [{\citenamefont {Wu}, \citenamefont {Liu},\ and\ \citenamefont
  {Liu}(2014)}]{Wu2014}%
  \BibitemOpen
  \bibfield  {author} {\bibinfo {author} {\bibfnamefont {J.}~\bibnamefont
  {Wu}}, \bibinfo {author} {\bibfnamefont {J.}~\bibnamefont {Liu}}, \ and\
  \bibinfo {author} {\bibfnamefont {X.~J.}\ \bibnamefont {Liu}},\ }\bibfield
  {title} {\enquote {\bibinfo {title} {{Topological spin texture in a quantum
  anomalous hall insulator}},}\ }\href {\doibase
  10.1103/PhysRevLett.113.136403} {\bibfield  {journal} {\bibinfo  {journal}
  {Phys. Rev. Lett.}\ }\textbf {\bibinfo {volume} {113}},\ \bibinfo {pages}
  {136403} (\bibinfo {year} {2014})},\ \Eprint {http://arxiv.org/abs/1401.0415}
  {arXiv:1401.0415} \BibitemShut {NoStop}%
\bibitem [{\citenamefont {Fert}, \citenamefont {Cros},\ and\ \citenamefont
  {Sampaio}(2013)}]{Fert2013}%
  \BibitemOpen
  \bibfield  {author} {\bibinfo {author} {\bibfnamefont {A.}~\bibnamefont
  {Fert}}, \bibinfo {author} {\bibfnamefont {V.}~\bibnamefont {Cros}}, \ and\
  \bibinfo {author} {\bibfnamefont {J.}~\bibnamefont {Sampaio}},\ }\bibfield
  {title} {\enquote {\bibinfo {title} {{Skyrmions on the track}},}\ }\href
  {\doibase 10.1038/nnano.2013.29} {\bibfield  {journal} {\bibinfo  {journal}
  {Nat. Nanotechnol.}\ }\textbf {\bibinfo {volume} {8}},\ \bibinfo {pages}
  {152--156} (\bibinfo {year} {2013})}\BibitemShut {NoStop}%
\bibitem [{\citenamefont {Burkov}(2016)}]{Burkov2016}%
  \BibitemOpen
  \bibfield  {author} {\bibinfo {author} {\bibfnamefont {A.~A.}\ \bibnamefont
  {Burkov}},\ }\bibfield  {title} {\enquote {\bibinfo {title} {{Topological
  semimetals}},}\ }\href {\doibase 10.1038/nmat4788} {\bibfield  {journal}
  {\bibinfo  {journal} {Nat. Mater.}\ }\textbf {\bibinfo {volume} {15}},\
  \bibinfo {pages} {1145--1148} (\bibinfo {year} {2016})},\ \Eprint
  {http://arxiv.org/abs/1610.07866} {arXiv:1610.07866} \BibitemShut {NoStop}%
\bibitem [{\citenamefont {Felser}\ and\ \citenamefont
  {Yan}(2016)}]{Felser2016}%
  \BibitemOpen
  \bibfield  {author} {\bibinfo {author} {\bibfnamefont {C.}~\bibnamefont
  {Felser}}\ and\ \bibinfo {author} {\bibfnamefont {B.}~\bibnamefont {Yan}},\
  }\bibfield  {title} {\enquote {\bibinfo {title} {{Weyl semimetals:
  Magnetically induced}},}\ }\href {\doibase 10.1038/nmat4741} {\bibfield
  {journal} {\bibinfo  {journal} {Nat. Mater.}\ }\textbf {\bibinfo {volume}
  {15}},\ \bibinfo {pages} {1149--1150} (\bibinfo {year} {2016})}\BibitemShut
  {NoStop}%
\bibitem [{\citenamefont {{\v{S}}mejkal}, \citenamefont {Jungwirth},\ and\
  \citenamefont {Sinova}(2017)}]{Smejkal2017}%
  \BibitemOpen
  \bibfield  {author} {\bibinfo {author} {\bibfnamefont {L.}~\bibnamefont
  {{\v{S}}mejkal}}, \bibinfo {author} {\bibfnamefont {T.}~\bibnamefont
  {Jungwirth}}, \ and\ \bibinfo {author} {\bibfnamefont {J.}~\bibnamefont
  {Sinova}},\ }\bibfield  {title} {\enquote {\bibinfo {title} {{Route Towards
  Dirac and Weyl Antiferromagnetic Spintronics}},}\ }\href {\doibase
  10.1002/pssr.201700044} {\bibfield  {journal} {\bibinfo  {journal} {Phys.
  Stat. Sol.}\ } (\bibinfo {year} {2017}),\ 10.1002/pssr.201700044},\ \Eprint
  {http://arxiv.org/abs/1702.07788} {arXiv:1702.07788} \BibitemShut {NoStop}%
\bibitem [{\citenamefont {Fan}\ \emph {et~al.}(2014)\citenamefont {Fan},
  \citenamefont {Upadhyaya}, \citenamefont {Kou}, \citenamefont {Lang},
  \citenamefont {Takei}, \citenamefont {Wang}, \citenamefont {Tang},
  \citenamefont {He}, \citenamefont {Chang}, \citenamefont {Montazeri},
  \citenamefont {Yu}, \citenamefont {Jiang}, \citenamefont {Nie}, \citenamefont
  {Schwartz}, \citenamefont {Tserkovnyak},\ and\ \citenamefont
  {Wang}}]{Fan2014a}%
  \BibitemOpen
  \bibfield  {author} {\bibinfo {author} {\bibfnamefont {Y.}~\bibnamefont
  {Fan}}, \bibinfo {author} {\bibfnamefont {P.}~\bibnamefont {Upadhyaya}},
  \bibinfo {author} {\bibfnamefont {X.}~\bibnamefont {Kou}}, \bibinfo {author}
  {\bibfnamefont {M.}~\bibnamefont {Lang}}, \bibinfo {author} {\bibfnamefont
  {S.}~\bibnamefont {Takei}}, \bibinfo {author} {\bibfnamefont
  {Z.}~\bibnamefont {Wang}}, \bibinfo {author} {\bibfnamefont {J.}~\bibnamefont
  {Tang}}, \bibinfo {author} {\bibfnamefont {L.}~\bibnamefont {He}}, \bibinfo
  {author} {\bibfnamefont {L.-T.}\ \bibnamefont {Chang}}, \bibinfo {author}
  {\bibfnamefont {M.}~\bibnamefont {Montazeri}}, \bibinfo {author}
  {\bibfnamefont {G.}~\bibnamefont {Yu}}, \bibinfo {author} {\bibfnamefont
  {W.}~\bibnamefont {Jiang}}, \bibinfo {author} {\bibfnamefont
  {T.}~\bibnamefont {Nie}}, \bibinfo {author} {\bibfnamefont {R.~N.}\
  \bibnamefont {Schwartz}}, \bibinfo {author} {\bibfnamefont {Y.}~\bibnamefont
  {Tserkovnyak}}, \ and\ \bibinfo {author} {\bibfnamefont {K.~L.}\ \bibnamefont
  {Wang}},\ }\bibfield  {title} {\enquote {\bibinfo {title} {{Magnetization
  switching through giant spin-orbit torque in a magnetically doped topological
  insulator heterostructure.}}}\ }\href {\doibase 10.1038/nmat3973} {\bibfield
  {journal} {\bibinfo  {journal} {Nat. Mater.}\ }\textbf {\bibinfo {volume}
  {13}},\ \bibinfo {pages} {699--704} (\bibinfo {year} {2014})}\BibitemShut
  {NoStop}%
\bibitem [{\citenamefont {Han}\ \emph {et~al.}(2017)\citenamefont {Han},
  \citenamefont {Richardella}, \citenamefont {Siddiqui}, \citenamefont
  {Finley}, \citenamefont {Samarth},\ and\ \citenamefont {Liu}}]{Han2017}%
  \BibitemOpen
  \bibfield  {author} {\bibinfo {author} {\bibfnamefont {J.}~\bibnamefont
  {Han}}, \bibinfo {author} {\bibfnamefont {A.}~\bibnamefont {Richardella}},
  \bibinfo {author} {\bibfnamefont {S.}~\bibnamefont {Siddiqui}}, \bibinfo
  {author} {\bibfnamefont {J.}~\bibnamefont {Finley}}, \bibinfo {author}
  {\bibfnamefont {N.}~\bibnamefont {Samarth}}, \ and\ \bibinfo {author}
  {\bibfnamefont {L.}~\bibnamefont {Liu}},\ }\bibfield  {title} {\enquote
  {\bibinfo {title} {{Room temperature spin-orbit torque switching induced by a
  topological insulator}},}\ }\href {http://arxiv.org/abs/1703.07470} {\
  (\bibinfo {year} {2017})},\ \Eprint {http://arxiv.org/abs/1703.07470}
  {arXiv:1703.07470} \BibitemShut {NoStop}%
\bibitem [{Note1()}]{Note1}%
  \BibitemOpen
  \bibinfo {note} {A recent report \cite {Katmis2016} of interfacial
  ferromagnetism persisting to room temperature at a insulating ferromagnet
  (EuS) /TI heterostructure is promising in this respect.}\BibitemShut {Stop}%
\bibitem [{\citenamefont {He}\ \emph {et~al.}(2016{\natexlab{a}})\citenamefont
  {He}, \citenamefont {Kou}, \citenamefont {Grutter}, \citenamefont {Yin},
  \citenamefont {Pan}, \citenamefont {Che}, \citenamefont {Liu}, \citenamefont
  {Nie}, \citenamefont {Zhang}, \citenamefont {Disseler}, \citenamefont
  {Kirby}, \citenamefont {{Ratcliff II}}, \citenamefont {Shao}, \citenamefont
  {Murata}, \citenamefont {Zhu}, \citenamefont {Yu}, \citenamefont {Fan},
  \citenamefont {Montazeri}, \citenamefont {Han}, \citenamefont {Borchers},\
  and\ \citenamefont {Wang}}]{He2016}%
  \BibitemOpen
  \bibfield  {author} {\bibinfo {author} {\bibfnamefont {Q.~L.}\ \bibnamefont
  {He}}, \bibinfo {author} {\bibfnamefont {X.}~\bibnamefont {Kou}}, \bibinfo
  {author} {\bibfnamefont {A.~J.}\ \bibnamefont {Grutter}}, \bibinfo {author}
  {\bibfnamefont {G.}~\bibnamefont {Yin}}, \bibinfo {author} {\bibfnamefont
  {L.}~\bibnamefont {Pan}}, \bibinfo {author} {\bibfnamefont {X.}~\bibnamefont
  {Che}}, \bibinfo {author} {\bibfnamefont {Y.}~\bibnamefont {Liu}}, \bibinfo
  {author} {\bibfnamefont {T.}~\bibnamefont {Nie}}, \bibinfo {author}
  {\bibfnamefont {B.}~\bibnamefont {Zhang}}, \bibinfo {author} {\bibfnamefont
  {S.~M.}\ \bibnamefont {Disseler}}, \bibinfo {author} {\bibfnamefont {B.~J.}\
  \bibnamefont {Kirby}}, \bibinfo {author} {\bibfnamefont {W.}~\bibnamefont
  {{Ratcliff II}}}, \bibinfo {author} {\bibfnamefont {Q.}~\bibnamefont {Shao}},
  \bibinfo {author} {\bibfnamefont {K.}~\bibnamefont {Murata}}, \bibinfo
  {author} {\bibfnamefont {X.}~\bibnamefont {Zhu}}, \bibinfo {author}
  {\bibfnamefont {G.}~\bibnamefont {Yu}}, \bibinfo {author} {\bibfnamefont
  {Y.}~\bibnamefont {Fan}}, \bibinfo {author} {\bibfnamefont {M.}~\bibnamefont
  {Montazeri}}, \bibinfo {author} {\bibfnamefont {X.}~\bibnamefont {Han}},
  \bibinfo {author} {\bibfnamefont {J.~A.}\ \bibnamefont {Borchers}}, \ and\
  \bibinfo {author} {\bibfnamefont {K.~L.}\ \bibnamefont {Wang}},\ }\bibfield
  {title} {\enquote {\bibinfo {title} {{Tailoring exchange couplings in
  magnetic topological-insulator/antiferromagnet heterostructures}},}\ }\href
  {\doibase 10.1038/nmat4783} {\bibfield  {journal} {\bibinfo  {journal} {Nat.
  Mater.}\ }\textbf {\bibinfo {volume} {16}},\ \bibinfo {pages} {94--100}
  (\bibinfo {year} {2016}{\natexlab{a}})}\BibitemShut {NoStop}%
\bibitem [{\citenamefont {Finley}\ and\ \citenamefont
  {Liu}(2016)}]{Finley2016}%
  \BibitemOpen
  \bibfield  {author} {\bibinfo {author} {\bibfnamefont {J.}~\bibnamefont
  {Finley}}\ and\ \bibinfo {author} {\bibfnamefont {L.}~\bibnamefont {Liu}},\
  }\bibfield  {title} {\enquote {\bibinfo {title} {{Spin-Orbit-Torque
  Efficiency in Compensated Ferrimagnetic Cobalt-Terbium Alloys}},}\ }\href
  {\doibase 10.1103/PhysRevApplied.6.054001} {\bibfield  {journal} {\bibinfo
  {journal} {Phys. Rev. Appl.}\ }\textbf {\bibinfo {volume} {6}},\ \bibinfo
  {pages} {054001} (\bibinfo {year} {2016})}\BibitemShut {NoStop}%
\bibitem [{\citenamefont {Katmis}\ \emph {et~al.}(2016)\citenamefont {Katmis},
  \citenamefont {Lauter}, \citenamefont {Nogueira}, \citenamefont {Assaf},
  \citenamefont {Jamer}, \citenamefont {Wei}, \citenamefont {Satpati},
  \citenamefont {Freeland}, \citenamefont {Eremin}, \citenamefont {Heiman},
  \citenamefont {Jarillo-Herrero},\ and\ \citenamefont {Moodera}}]{Katmis2016}%
  \BibitemOpen
  \bibfield  {author} {\bibinfo {author} {\bibfnamefont {F.}~\bibnamefont
  {Katmis}}, \bibinfo {author} {\bibfnamefont {V.}~\bibnamefont {Lauter}},
  \bibinfo {author} {\bibfnamefont {F.~S.}\ \bibnamefont {Nogueira}}, \bibinfo
  {author} {\bibfnamefont {B.~A.}\ \bibnamefont {Assaf}}, \bibinfo {author}
  {\bibfnamefont {M.~E.}\ \bibnamefont {Jamer}}, \bibinfo {author}
  {\bibfnamefont {P.}~\bibnamefont {Wei}}, \bibinfo {author} {\bibfnamefont
  {B.}~\bibnamefont {Satpati}}, \bibinfo {author} {\bibfnamefont {J.~W.}\
  \bibnamefont {Freeland}}, \bibinfo {author} {\bibfnamefont {I.}~\bibnamefont
  {Eremin}}, \bibinfo {author} {\bibfnamefont {D.}~\bibnamefont {Heiman}},
  \bibinfo {author} {\bibfnamefont {P.}~\bibnamefont {Jarillo-Herrero}}, \ and\
  \bibinfo {author} {\bibfnamefont {J.~S.}\ \bibnamefont {Moodera}},\
  }\bibfield  {title} {\enquote {\bibinfo {title} {{A high-temperature
  ferromagnetic topological insulating phase by proximity coupling}},}\ }\href
  {\doibase 10.1038/nature17635} {\bibfield  {journal} {\bibinfo  {journal}
  {Nature}\ }\textbf {\bibinfo {volume} {533}},\ \bibinfo {pages} {513--516}
  (\bibinfo {year} {2016})}\BibitemShut {NoStop}%
\bibitem [{\citenamefont {Park}\ \emph {et~al.}(2011)\citenamefont {Park},
  \citenamefont {Lee}, \citenamefont {Wolff-Fabris}, \citenamefont {Koh},
  \citenamefont {Eom}, \citenamefont {Kim}, \citenamefont {Farhan},
  \citenamefont {Jo}, \citenamefont {Kim}, \citenamefont {Shim},\ and\
  \citenamefont {Kim}}]{Park2011a}%
  \BibitemOpen
  \bibfield  {author} {\bibinfo {author} {\bibfnamefont {J.}~\bibnamefont
  {Park}}, \bibinfo {author} {\bibfnamefont {G.}~\bibnamefont {Lee}}, \bibinfo
  {author} {\bibfnamefont {F.}~\bibnamefont {Wolff-Fabris}}, \bibinfo {author}
  {\bibfnamefont {Y.~Y.}\ \bibnamefont {Koh}}, \bibinfo {author} {\bibfnamefont
  {M.~J.}\ \bibnamefont {Eom}}, \bibinfo {author} {\bibfnamefont {Y.~K.}\
  \bibnamefont {Kim}}, \bibinfo {author} {\bibfnamefont {M.~A.}\ \bibnamefont
  {Farhan}}, \bibinfo {author} {\bibfnamefont {Y.~J.}\ \bibnamefont {Jo}},
  \bibinfo {author} {\bibfnamefont {C.}~\bibnamefont {Kim}}, \bibinfo {author}
  {\bibfnamefont {J.~H.}\ \bibnamefont {Shim}}, \ and\ \bibinfo {author}
  {\bibfnamefont {J.~S.}\ \bibnamefont {Kim}},\ }\bibfield  {title} {\enquote
  {\bibinfo {title} {{Anisotropic Dirac Fermions in a Bi Square Net of
  SrMnBi2}},}\ }\href {\doibase 10.1103/PhysRevLett.107.126402} {\bibfield
  {journal} {\bibinfo  {journal} {Phys. Rev. Lett.}\ }\textbf {\bibinfo
  {volume} {107}},\ \bibinfo {pages} {126402} (\bibinfo {year} {2011})},\
  \Eprint {http://arxiv.org/abs/1104.5138} {arXiv:1104.5138} \BibitemShut
  {NoStop}%
\bibitem [{\citenamefont {Wang}\ \emph {et~al.}(2011)\citenamefont {Wang},
  \citenamefont {Graf}, \citenamefont {Lei}, \citenamefont {Tozer},\ and\
  \citenamefont {Petrovic}}]{Wang2011e}%
  \BibitemOpen
  \bibfield  {author} {\bibinfo {author} {\bibfnamefont {K.}~\bibnamefont
  {Wang}}, \bibinfo {author} {\bibfnamefont {D.}~\bibnamefont {Graf}}, \bibinfo
  {author} {\bibfnamefont {H.}~\bibnamefont {Lei}}, \bibinfo {author}
  {\bibfnamefont {S.~W.}\ \bibnamefont {Tozer}}, \ and\ \bibinfo {author}
  {\bibfnamefont {C.}~\bibnamefont {Petrovic}},\ }\bibfield  {title} {\enquote
  {\bibinfo {title} {{Quantum transport of two-dimensional Dirac fermions in
  SrMnBi2}},}\ }\href {\doibase 10.1103/PhysRevB.84.220401} {\bibfield
  {journal} {\bibinfo  {journal} {Phys. Rev. B - Condens. Matter Mater. Phys.}\
  }\textbf {\bibinfo {volume} {84}},\ \bibinfo {pages} {220401(R)} (\bibinfo
  {year} {2011})},\ \Eprint {http://arxiv.org/abs/1204.1049v1}
  {arXiv:1204.1049v1} \BibitemShut {NoStop}%
\bibitem [{\citenamefont {Masuda}\ \emph {et~al.}(2016)\citenamefont {Masuda},
  \citenamefont {Sakai}, \citenamefont {Tokunaga}, \citenamefont {Yamasaki},
  \citenamefont {Miyake}, \citenamefont {Shiogai}, \citenamefont {Nakamura},
  \citenamefont {Awaji}, \citenamefont {Tsukazaki}, \citenamefont {Nakao},
  \citenamefont {Murakami}, \citenamefont {Arima}, \citenamefont {Tokura},\
  and\ \citenamefont {Ishiwata}}]{Masuda2016}%
  \BibitemOpen
  \bibfield  {author} {\bibinfo {author} {\bibfnamefont {H.}~\bibnamefont
  {Masuda}}, \bibinfo {author} {\bibfnamefont {H.}~\bibnamefont {Sakai}},
  \bibinfo {author} {\bibfnamefont {M.}~\bibnamefont {Tokunaga}}, \bibinfo
  {author} {\bibfnamefont {Y.}~\bibnamefont {Yamasaki}}, \bibinfo {author}
  {\bibfnamefont {A.}~\bibnamefont {Miyake}}, \bibinfo {author} {\bibfnamefont
  {J.}~\bibnamefont {Shiogai}}, \bibinfo {author} {\bibfnamefont
  {S.}~\bibnamefont {Nakamura}}, \bibinfo {author} {\bibfnamefont
  {S.}~\bibnamefont {Awaji}}, \bibinfo {author} {\bibfnamefont
  {A.}~\bibnamefont {Tsukazaki}}, \bibinfo {author} {\bibfnamefont
  {H.}~\bibnamefont {Nakao}}, \bibinfo {author} {\bibfnamefont
  {Y.}~\bibnamefont {Murakami}}, \bibinfo {author} {\bibfnamefont {T.-h.}\
  \bibnamefont {Arima}}, \bibinfo {author} {\bibfnamefont {Y.}~\bibnamefont
  {Tokura}}, \ and\ \bibinfo {author} {\bibfnamefont {S.}~\bibnamefont
  {Ishiwata}},\ }\bibfield  {title} {\enquote {\bibinfo {title} {{Quantum Hall
  effect in a bulk antiferromagnet EuMnBi2 with magnetically confined
  two-dimensional Dirac fermions}},}\ }\href {\doibase 10.1126/sciadv.1501117}
  {\bibfield  {journal} {\bibinfo  {journal} {Sci. Adv.}\ }\textbf {\bibinfo
  {volume} {2}},\ \bibinfo {pages} {e1501117} (\bibinfo {year}
  {2016})}\BibitemShut {NoStop}%
\bibitem [{\citenamefont {Mong}, \citenamefont {Essin},\ and\ \citenamefont
  {Moore}(2010)}]{Mong2010}%
  \BibitemOpen
  \bibfield  {author} {\bibinfo {author} {\bibfnamefont {R.~S.~K.}\
  \bibnamefont {Mong}}, \bibinfo {author} {\bibfnamefont {A.~M.}\ \bibnamefont
  {Essin}}, \ and\ \bibinfo {author} {\bibfnamefont {J.~E.}\ \bibnamefont
  {Moore}},\ }\bibfield  {title} {\enquote {\bibinfo {title}
  {{Antiferromagnetic topological insulators}},}\ }\href {\doibase
  10.1103/PhysRevB.81.245209} {\bibfield  {journal} {\bibinfo  {journal} {Phys.
  Rev. B}\ }\textbf {\bibinfo {volume} {81}},\ \bibinfo {pages} {245209}
  (\bibinfo {year} {2010})},\ \Eprint {http://arxiv.org/abs/1004.1403}
  {arXiv:1004.1403} \BibitemShut {NoStop}%
\bibitem [{\citenamefont {Liu}\ \emph {et~al.}(2011)\citenamefont {Liu},
  \citenamefont {Lee}, \citenamefont {Kondo}, \citenamefont {Mun},
  \citenamefont {Caudle}, \citenamefont {Harmon}, \citenamefont {Bud},
  \citenamefont {Canfield},\ and\ \citenamefont {Kaminski}}]{Liu2011f}%
  \BibitemOpen
  \bibfield  {author} {\bibinfo {author} {\bibfnamefont {C.}~\bibnamefont
  {Liu}}, \bibinfo {author} {\bibfnamefont {Y.}~\bibnamefont {Lee}}, \bibinfo
  {author} {\bibfnamefont {T.}~\bibnamefont {Kondo}}, \bibinfo {author}
  {\bibfnamefont {E.~D.}\ \bibnamefont {Mun}}, \bibinfo {author} {\bibfnamefont
  {M.}~\bibnamefont {Caudle}}, \bibinfo {author} {\bibfnamefont {B.~N.}\
  \bibnamefont {Harmon}}, \bibinfo {author} {\bibfnamefont {S.~L.}\
  \bibnamefont {Bud}}, \bibinfo {author} {\bibfnamefont {P.~C.}\ \bibnamefont
  {Canfield}}, \ and\ \bibinfo {author} {\bibfnamefont {A.}~\bibnamefont
  {Kaminski}},\ }\bibfield  {title} {\enquote {\bibinfo {title} {{Metallic
  surface electronic state in half-Heusler compounds RPtBi (R = Lu , Dy ,
  Gd)}},}\ }\href {\doibase 10.1103/PhysRevB.83.205133} {\ \textbf {\bibinfo
  {volume} {83}},\ \bibinfo {pages} {205133 (2011)} (\bibinfo {year}
  {2011})}\BibitemShut {NoStop}%
\bibitem [{\citenamefont {Wang}\ \emph
  {et~al.}(2016{\natexlab{b}})\citenamefont {Wang}, \citenamefont {Zhang},
  \citenamefont {Liu}, \citenamefont {Liu}, \citenamefont {Tang}, \citenamefont
  {Song}, \citenamefont {Zhong}, \citenamefont {Peng}, \citenamefont {Li},
  \citenamefont {Nie}, \citenamefont {Wang}, \citenamefont {Zhou},
  \citenamefont {Ma}, \citenamefont {Xue},\ and\ \citenamefont
  {Liu}}]{Wang2016e}%
  \BibitemOpen
  \bibfield  {author} {\bibinfo {author} {\bibfnamefont {Z.~F.}\ \bibnamefont
  {Wang}}, \bibinfo {author} {\bibfnamefont {H.}~\bibnamefont {Zhang}},
  \bibinfo {author} {\bibfnamefont {D.}~\bibnamefont {Liu}}, \bibinfo {author}
  {\bibfnamefont {C.}~\bibnamefont {Liu}}, \bibinfo {author} {\bibfnamefont
  {C.}~\bibnamefont {Tang}}, \bibinfo {author} {\bibfnamefont {C.}~\bibnamefont
  {Song}}, \bibinfo {author} {\bibfnamefont {Y.}~\bibnamefont {Zhong}},
  \bibinfo {author} {\bibfnamefont {J.}~\bibnamefont {Peng}}, \bibinfo {author}
  {\bibfnamefont {F.}~\bibnamefont {Li}}, \bibinfo {author} {\bibfnamefont
  {C.}~\bibnamefont {Nie}}, \bibinfo {author} {\bibfnamefont {L.}~\bibnamefont
  {Wang}}, \bibinfo {author} {\bibfnamefont {X.~J.}\ \bibnamefont {Zhou}},
  \bibinfo {author} {\bibfnamefont {X.}~\bibnamefont {Ma}}, \bibinfo {author}
  {\bibfnamefont {Q.~K.}\ \bibnamefont {Xue}}, \ and\ \bibinfo {author}
  {\bibfnamefont {F.}~\bibnamefont {Liu}},\ }\bibfield  {title} {\enquote
  {\bibinfo {title} {{Topological edge states in a high-temperature
  superconductor FeSe/SrTiO3(001) film}},}\ }\href {\doibase 10.1038/nmat4686}
  {\bibfield  {journal} {\bibinfo  {journal} {Nat. Mater.}\ }\textbf {\bibinfo
  {volume} {15}},\ \bibinfo {pages} {968} (\bibinfo {year}
  {2016}{\natexlab{b}})}\BibitemShut {NoStop}%
\bibitem [{\citenamefont {Tsai}\ and\ \citenamefont {Lin}(2016)}]{Tsai2016}%
  \BibitemOpen
  \bibfield  {author} {\bibinfo {author} {\bibfnamefont {W.-F.}\ \bibnamefont
  {Tsai}}\ and\ \bibinfo {author} {\bibfnamefont {H.}~\bibnamefont {Lin}},\
  }\bibfield  {title} {\enquote {\bibinfo {title} {{Topological insulators and
  superconductivity: The integrity of two sides}},}\ }\href {\doibase
  10.1038/nmat4700} {\bibfield  {journal} {\bibinfo  {journal} {Nat. Mater.}\ }
  (\bibinfo {year} {2016}),\ 10.1038/nmat4700}\BibitemShut {NoStop}%
\bibitem [{\citenamefont {Fu}\ and\ \citenamefont {Kane}(2008)}]{Fu2008}%
  \BibitemOpen
  \bibfield  {author} {\bibinfo {author} {\bibfnamefont {L.}~\bibnamefont
  {Fu}}\ and\ \bibinfo {author} {\bibfnamefont {C.~L.}\ \bibnamefont {Kane}},\
  }\bibfield  {title} {\enquote {\bibinfo {title} {{Superconducting Proximity
  Effect and Majorana Fermions at the Surface of a Topological Insulator}},}\
  }\href {\doibase 10.1103/PhysRevLett.100.096407} {\bibfield  {journal}
  {\bibinfo  {journal} {Phys. Rev. Lett.}\ }\textbf {\bibinfo {volume} {100}},\
  \bibinfo {pages} {096407} (\bibinfo {year} {2008})}\BibitemShut {NoStop}%
\bibitem [{\citenamefont {Niu}\ \emph {et~al.}(2017)\citenamefont {Niu},
  \citenamefont {Hanke}, \citenamefont {Buhl}, \citenamefont {Bihlmayer},
  \citenamefont {Wortmann}, \citenamefont {Bl{\"{u}}gel},\ and\ \citenamefont
  {Mokrousov}}]{Niu2017}%
  \BibitemOpen
  \bibfield  {author} {\bibinfo {author} {\bibfnamefont {C.}~\bibnamefont
  {Niu}}, \bibinfo {author} {\bibfnamefont {J.~P.}\ \bibnamefont {Hanke}},
  \bibinfo {author} {\bibfnamefont {P.~M.}\ \bibnamefont {Buhl}}, \bibinfo
  {author} {\bibfnamefont {G.}~\bibnamefont {Bihlmayer}}, \bibinfo {author}
  {\bibfnamefont {D.}~\bibnamefont {Wortmann}}, \bibinfo {author}
  {\bibfnamefont {S.}~\bibnamefont {Bl{\"{u}}gel}}, \ and\ \bibinfo {author}
  {\bibfnamefont {Y.}~\bibnamefont {Mokrousov}},\ }\bibfield  {title} {\enquote
  {\bibinfo {title} {{Quantum spin Hall effect and topological phase
  transitions in honeycomb antiferromagnets}},}\ }\href
  {http://arxiv.org/abs/1705.07035} {\ ,\ \bibinfo {pages} {1--5} (\bibinfo
  {year} {2017})},\ \Eprint {http://arxiv.org/abs/1705.07035}
  {arXiv:1705.07035} \BibitemShut {NoStop}%
\bibitem [{\citenamefont {Armitage}, \citenamefont {Mele},\ and\ \citenamefont
  {Vishwanath}(2017)}]{Armitage2017}%
  \BibitemOpen
  \bibfield  {author} {\bibinfo {author} {\bibfnamefont {N.~P.}\ \bibnamefont
  {Armitage}}, \bibinfo {author} {\bibfnamefont {E.~J.}\ \bibnamefont {Mele}},
  \ and\ \bibinfo {author} {\bibfnamefont {A.}~\bibnamefont {Vishwanath}},\
  }\bibfield  {title} {\enquote {\bibinfo {title} {{Weyl and Dirac Semimetals
  in Three Dimensional Solids ∗}},}\ }\href@noop {} {\  (\bibinfo {year}
  {2017})},\ \Eprint {http://arxiv.org/abs/1705.01111v1} {arXiv:1705.01111v1}
  \BibitemShut {NoStop}%
\bibitem [{\citenamefont {Jia}, \citenamefont {Xu},\ and\ \citenamefont
  {Hasan}(2016)}]{Jia2016}%
  \BibitemOpen
  \bibfield  {author} {\bibinfo {author} {\bibfnamefont {S.}~\bibnamefont
  {Jia}}, \bibinfo {author} {\bibfnamefont {S.-Y.}\ \bibnamefont {Xu}}, \ and\
  \bibinfo {author} {\bibfnamefont {M.~Z.}\ \bibnamefont {Hasan}},\ }\bibfield
  {title} {\enquote {\bibinfo {title} {{Weyl semimetals, Fermi arcs and chiral
  anomalies}},}\ }\href {\doibase 10.1038/nmat4787} {\bibfield  {journal}
  {\bibinfo  {journal} {Nat. Mater.}\ }\textbf {\bibinfo {volume} {15}},\
  \bibinfo {pages} {1140--1144} (\bibinfo {year} {2016})}\BibitemShut {NoStop}%
\bibitem [{\citenamefont {Hirschberger}\ \emph {et~al.}(2016)\citenamefont
  {Hirschberger}, \citenamefont {Kushwaha}, \citenamefont {Wang}, \citenamefont
  {Gibson}, \citenamefont {Liang}, \citenamefont {Belvin}, \citenamefont
  {Bernevig}, \citenamefont {Cava},\ and\ \citenamefont
  {Ong}}]{Hirschberger2016}%
  \BibitemOpen
  \bibfield  {author} {\bibinfo {author} {\bibfnamefont {M.}~\bibnamefont
  {Hirschberger}}, \bibinfo {author} {\bibfnamefont {S.}~\bibnamefont
  {Kushwaha}}, \bibinfo {author} {\bibfnamefont {Z.}~\bibnamefont {Wang}},
  \bibinfo {author} {\bibfnamefont {Q.}~\bibnamefont {Gibson}}, \bibinfo
  {author} {\bibfnamefont {S.}~\bibnamefont {Liang}}, \bibinfo {author}
  {\bibfnamefont {C.~A.}\ \bibnamefont {Belvin}}, \bibinfo {author}
  {\bibfnamefont {B.~A.}\ \bibnamefont {Bernevig}}, \bibinfo {author}
  {\bibfnamefont {R.~J.}\ \bibnamefont {Cava}}, \ and\ \bibinfo {author}
  {\bibfnamefont {N.~P.}\ \bibnamefont {Ong}},\ }\bibfield  {title} {\enquote
  {\bibinfo {title} {{The chiral anomaly and thermopower of Weyl fermions in
  the half-Heusler GdPtBi}},}\ }\href {\doibase 10.1038/nmat4684} {\bibfield
  {journal} {\bibinfo  {journal} {Nat. Mater.}\ }\textbf {\bibinfo {volume}
  {15}},\ \bibinfo {pages} {1161--1165} (\bibinfo {year} {2016})},\ \Eprint
  {http://arxiv.org/abs/1602.07219} {arXiv:1602.07219} \BibitemShut {NoStop}%
\bibitem [{\citenamefont {Ali}\ \emph {et~al.}(2014)\citenamefont {Ali},
  \citenamefont {Xiong}, \citenamefont {Flynn}, \citenamefont {Tao},
  \citenamefont {Gibson}, \citenamefont {Schoop}, \citenamefont {Liang},
  \citenamefont {Haldolaarachchige}, \citenamefont {Hirschberger},
  \citenamefont {Ong},\ and\ \citenamefont {Cava}}]{Ali2014}%
  \BibitemOpen
  \bibfield  {author} {\bibinfo {author} {\bibfnamefont {M.~N.}\ \bibnamefont
  {Ali}}, \bibinfo {author} {\bibfnamefont {J.}~\bibnamefont {Xiong}}, \bibinfo
  {author} {\bibfnamefont {S.}~\bibnamefont {Flynn}}, \bibinfo {author}
  {\bibfnamefont {J.}~\bibnamefont {Tao}}, \bibinfo {author} {\bibfnamefont
  {Q.~D.}\ \bibnamefont {Gibson}}, \bibinfo {author} {\bibfnamefont {L.~M.}\
  \bibnamefont {Schoop}}, \bibinfo {author} {\bibfnamefont {T.}~\bibnamefont
  {Liang}}, \bibinfo {author} {\bibfnamefont {N.}~\bibnamefont
  {Haldolaarachchige}}, \bibinfo {author} {\bibfnamefont {M.}~\bibnamefont
  {Hirschberger}}, \bibinfo {author} {\bibfnamefont {N.~P.}\ \bibnamefont
  {Ong}}, \ and\ \bibinfo {author} {\bibfnamefont {R.~J.}\ \bibnamefont
  {Cava}},\ }\bibfield  {title} {\enquote {\bibinfo {title} {{Large,
  non-saturating magnetoresistance in WTe2.}}}\ }\href {\doibase
  10.1038/nature13763} {\bibfield  {journal} {\bibinfo  {journal} {Nature}\
  }\textbf {\bibinfo {volume} {514}},\ \bibinfo {pages} {205--208} (\bibinfo
  {year} {2014})},\ \Eprint {http://arxiv.org/abs/1405.0973} {arXiv:1405.0973}
  \BibitemShut {NoStop}%
\bibitem [{\citenamefont {Soluyanov}\ \emph {et~al.}(2015)\citenamefont
  {Soluyanov}, \citenamefont {Gresch}, \citenamefont {Wang}, \citenamefont
  {Wu}, \citenamefont {Troyer}, \citenamefont {Dai},\ and\ \citenamefont
  {Bernevig}}]{Soluyanov2015}%
  \BibitemOpen
  \bibfield  {author} {\bibinfo {author} {\bibfnamefont {A.~A.}\ \bibnamefont
  {Soluyanov}}, \bibinfo {author} {\bibfnamefont {D.}~\bibnamefont {Gresch}},
  \bibinfo {author} {\bibfnamefont {Z.}~\bibnamefont {Wang}}, \bibinfo {author}
  {\bibfnamefont {Q.}~\bibnamefont {Wu}}, \bibinfo {author} {\bibfnamefont
  {M.}~\bibnamefont {Troyer}}, \bibinfo {author} {\bibfnamefont
  {X.}~\bibnamefont {Dai}}, \ and\ \bibinfo {author} {\bibfnamefont {B.~A.}\
  \bibnamefont {Bernevig}},\ }\bibfield  {title} {\enquote {\bibinfo {title}
  {{Type-II Weyl semimetals}},}\ }\href {\doibase 10.1038/nature15768}
  {\bibfield  {journal} {\bibinfo  {journal} {Nature}\ }\textbf {\bibinfo
  {volume} {527}},\ \bibinfo {pages} {495--498} (\bibinfo {year}
  {2015})}\BibitemShut {NoStop}%
\bibitem [{\citenamefont {Pletikosic}\ \emph {et~al.}(2014)\citenamefont
  {Pletikosic}, \citenamefont {Ali}, \citenamefont {Fedorov}, \citenamefont
  {Cava},\ and\ \citenamefont {Valla}}]{Pletikosic2014}%
  \BibitemOpen
  \bibfield  {author} {\bibinfo {author} {\bibfnamefont {I.}~\bibnamefont
  {Pletikosic}}, \bibinfo {author} {\bibfnamefont {M.~N.}\ \bibnamefont {Ali}},
  \bibinfo {author} {\bibfnamefont {A.~V.}\ \bibnamefont {Fedorov}}, \bibinfo
  {author} {\bibfnamefont {R.~J.}\ \bibnamefont {Cava}}, \ and\ \bibinfo
  {author} {\bibfnamefont {T.}~\bibnamefont {Valla}},\ }\bibfield  {title}
  {\enquote {\bibinfo {title} {{Electronic structure basis for the
  extraordinary magnetoresistance in WTe2}},}\ }\href {\doibase
  10.1103/PhysRevLett.113.216601} {\bibfield  {journal} {\bibinfo  {journal}
  {Phys. Rev. Lett.}\ }\textbf {\bibinfo {volume} {113}},\ \bibinfo {pages}
  {216601} (\bibinfo {year} {2014})},\ \Eprint
  {http://arxiv.org/abs/1407.3576v1} {arXiv:1407.3576v1} \BibitemShut {NoStop}%
\bibitem [{\citenamefont {Yang}\ and\ \citenamefont
  {Nagaosa}(2014)}]{Yang2014a}%
  \BibitemOpen
  \bibfield  {author} {\bibinfo {author} {\bibfnamefont {B.-j.}\ \bibnamefont
  {Yang}}\ and\ \bibinfo {author} {\bibfnamefont {N.}~\bibnamefont {Nagaosa}},\
  }\bibfield  {title} {\enquote {\bibinfo {title} {{Classification of stable
  three-dimensional Dirac semimetals with nontrivial topology}},}\ }\href
  {\doibase 10.1038/ncomms5898} {\bibfield  {journal} {\bibinfo  {journal}
  {Nat. Commun.}\ }\textbf {\bibinfo {volume} {5}},\ \bibinfo {pages} {4898}
  (\bibinfo {year} {2014})},\ \Eprint {http://arxiv.org/abs/1404.0754v1}
  {arXiv:1404.0754v1} \BibitemShut {NoStop}%
\bibitem [{\citenamefont {Kargarian}, \citenamefont {Randeria},\ and\
  \citenamefont {Lu}(2016)}]{Kargarian2016}%
  \BibitemOpen
  \bibfield  {author} {\bibinfo {author} {\bibfnamefont {M.}~\bibnamefont
  {Kargarian}}, \bibinfo {author} {\bibfnamefont {M.}~\bibnamefont {Randeria}},
  \ and\ \bibinfo {author} {\bibfnamefont {Y.-M.}\ \bibnamefont {Lu}},\
  }\bibfield  {title} {\enquote {\bibinfo {title} {{Are the surface Fermi arcs
  in Dirac semimetals topologically protected?}}}\ }\href {\doibase
  10.1073/pnas.1524787113} {\bibfield  {journal} {\bibinfo  {journal} {Proc.
  Natl. Acad. Sci.}\ }\textbf {\bibinfo {volume} {113}},\ \bibinfo {pages}
  {8648--8652} (\bibinfo {year} {2016})}\BibitemShut {NoStop}%
\bibitem [{\citenamefont {Tang}\ \emph {et~al.}(2016)\citenamefont {Tang},
  \citenamefont {Zhou}, \citenamefont {Xu},\ and\ \citenamefont
  {Zhang}}]{Tang2016}%
  \BibitemOpen
  \bibfield  {author} {\bibinfo {author} {\bibfnamefont {P.}~\bibnamefont
  {Tang}}, \bibinfo {author} {\bibfnamefont {Q.}~\bibnamefont {Zhou}}, \bibinfo
  {author} {\bibfnamefont {G.}~\bibnamefont {Xu}}, \ and\ \bibinfo {author}
  {\bibfnamefont {S.-C.}\ \bibnamefont {Zhang}},\ }\bibfield  {title} {\enquote
  {\bibinfo {title} {{Dirac fermions in an antiferromagnetic semimetal}},}\
  }\href {\doibase 10.1038/NPHYS3839} {\bibfield  {journal} {\bibinfo
  {journal} {Nat. Phys.}\ }\textbf {\bibinfo {volume} {12}},\ \bibinfo {pages}
  {1100--1104} (\bibinfo {year} {2016})}\BibitemShut {NoStop}%
\bibitem [{\citenamefont {{\v{S}}mejkal}\ \emph {et~al.}(2017)\citenamefont
  {{\v{S}}mejkal}, \citenamefont {{\v{Z}}elezn{\'{y}}}, \citenamefont
  {Sinova},\ and\ \citenamefont {Jungwirth}}]{Smejkal2016}%
  \BibitemOpen
  \bibfield  {author} {\bibinfo {author} {\bibfnamefont {L.}~\bibnamefont
  {{\v{S}}mejkal}}, \bibinfo {author} {\bibfnamefont {J.}~\bibnamefont
  {{\v{Z}}elezn{\'{y}}}}, \bibinfo {author} {\bibfnamefont {J.}~\bibnamefont
  {Sinova}}, \ and\ \bibinfo {author} {\bibfnamefont {T.}~\bibnamefont
  {Jungwirth}},\ }\bibfield  {title} {\enquote {\bibinfo {title} {{Electric
  Control of Dirac Quasiparticles by Spin-Orbit Torque in an
  Antiferromagnet}},}\ }\href {\doibase 10.1103/PhysRevLett.118.106402}
  {\bibfield  {journal} {\bibinfo  {journal} {Phys. Rev. Lett.}\ }\textbf
  {\bibinfo {volume} {118}},\ \bibinfo {pages} {106402} (\bibinfo {year}
  {2017})}\BibitemShut {NoStop}%
\bibitem [{\citenamefont {Young}\ and\ \citenamefont
  {Wieder}(2017)}]{Young2016}%
  \BibitemOpen
  \bibfield  {author} {\bibinfo {author} {\bibfnamefont {S.~M.}\ \bibnamefont
  {Young}}\ and\ \bibinfo {author} {\bibfnamefont {B.~J.}\ \bibnamefont
  {Wieder}},\ }\bibfield  {title} {\enquote {\bibinfo {title}
  {{Filling-enforced Magnetic Dirac Semimetals in Two Dimensions}},}\ }\href
  {http://arxiv.org/abs/1609.06738} {\bibfield  {journal} {\bibinfo  {journal}
  {Phys. Rev. Lett.}\ }\textbf {\bibinfo {volume} {118}},\ \bibinfo {pages}
  {186401} (\bibinfo {year} {2017})},\ \Eprint
  {http://arxiv.org/abs/1609.06738} {arXiv:1609.06738} \BibitemShut {NoStop}%
\bibitem [{\citenamefont {Wadley}\ \emph {et~al.}(2016)\citenamefont {Wadley},
  \citenamefont {Howells}, \citenamefont {Zelezny}, \citenamefont {Andrews},
  \citenamefont {Hills}, \citenamefont {Campion}, \citenamefont {Novak},
  \citenamefont {Olejn{\'{i}}k}, \citenamefont {Maccherozzi}, \citenamefont
  {Dhesi}, \citenamefont {Martin}, \citenamefont {Wagner}, \citenamefont
  {Wunderlich}, \citenamefont {Freimuth}, \citenamefont {Mokrousov},
  \citenamefont {Kunes}, \citenamefont {Chauhan}, \citenamefont {Grzybowski},
  \citenamefont {Rushforth}, \citenamefont {Edmonds}, \citenamefont
  {Gallagher},\ and\ \citenamefont {Jungwirth}}]{Wadley2016}%
  \BibitemOpen
  \bibfield  {author} {\bibinfo {author} {\bibfnamefont {P.}~\bibnamefont
  {Wadley}}, \bibinfo {author} {\bibfnamefont {B.}~\bibnamefont {Howells}},
  \bibinfo {author} {\bibfnamefont {J.}~\bibnamefont {Zelezny}}, \bibinfo
  {author} {\bibfnamefont {C.}~\bibnamefont {Andrews}}, \bibinfo {author}
  {\bibfnamefont {V.}~\bibnamefont {Hills}}, \bibinfo {author} {\bibfnamefont
  {R.~P.}\ \bibnamefont {Campion}}, \bibinfo {author} {\bibfnamefont
  {V.}~\bibnamefont {Novak}}, \bibinfo {author} {\bibfnamefont
  {K.}~\bibnamefont {Olejn{\'{i}}k}}, \bibinfo {author} {\bibfnamefont
  {F.}~\bibnamefont {Maccherozzi}}, \bibinfo {author} {\bibfnamefont {S.~S.}\
  \bibnamefont {Dhesi}}, \bibinfo {author} {\bibfnamefont {S.~Y.}\ \bibnamefont
  {Martin}}, \bibinfo {author} {\bibfnamefont {T.}~\bibnamefont {Wagner}},
  \bibinfo {author} {\bibfnamefont {J.}~\bibnamefont {Wunderlich}}, \bibinfo
  {author} {\bibfnamefont {F.}~\bibnamefont {Freimuth}}, \bibinfo {author}
  {\bibfnamefont {Y.}~\bibnamefont {Mokrousov}}, \bibinfo {author}
  {\bibfnamefont {J.}~\bibnamefont {Kunes}}, \bibinfo {author} {\bibfnamefont
  {J.~S.}\ \bibnamefont {Chauhan}}, \bibinfo {author} {\bibfnamefont {M.~J.}\
  \bibnamefont {Grzybowski}}, \bibinfo {author} {\bibfnamefont {A.~W.}\
  \bibnamefont {Rushforth}}, \bibinfo {author} {\bibfnamefont {K.~W.}\
  \bibnamefont {Edmonds}}, \bibinfo {author} {\bibfnamefont {B.~L.}\
  \bibnamefont {Gallagher}}, \ and\ \bibinfo {author} {\bibfnamefont
  {T.}~\bibnamefont {Jungwirth}},\ }\bibfield  {title} {\enquote {\bibinfo
  {title} {{Electrical switching of an antiferromagnet}},}\ }\href {\doibase
  10.1126/science.aab1031} {\bibfield  {journal} {\bibinfo  {journal}
  {Science}\ }\textbf {\bibinfo {volume} {351}},\ \bibinfo {pages} {587--590}
  (\bibinfo {year} {2016})}\BibitemShut {NoStop}%
\bibitem [{\citenamefont {Tian}\ \emph {et~al.}(2015)\citenamefont {Tian},
  \citenamefont {Kohama}, \citenamefont {Tomita}, \citenamefont {Ishizuka},
  \citenamefont {Hsieh}, \citenamefont {Ishikawa}, \citenamefont {Kindo},
  \citenamefont {Balents},\ and\ \citenamefont {Nakatsuji}}]{Tian2015}%
  \BibitemOpen
  \bibfield  {author} {\bibinfo {author} {\bibfnamefont {Z.}~\bibnamefont
  {Tian}}, \bibinfo {author} {\bibfnamefont {Y.}~\bibnamefont {Kohama}},
  \bibinfo {author} {\bibfnamefont {T.}~\bibnamefont {Tomita}}, \bibinfo
  {author} {\bibfnamefont {H.}~\bibnamefont {Ishizuka}}, \bibinfo {author}
  {\bibfnamefont {T.~H.}\ \bibnamefont {Hsieh}}, \bibinfo {author}
  {\bibfnamefont {J.~J.}\ \bibnamefont {Ishikawa}}, \bibinfo {author}
  {\bibfnamefont {K.}~\bibnamefont {Kindo}}, \bibinfo {author} {\bibfnamefont
  {L.}~\bibnamefont {Balents}}, \ and\ \bibinfo {author} {\bibfnamefont
  {S.}~\bibnamefont {Nakatsuji}},\ }\bibfield  {title} {\enquote {\bibinfo
  {title} {{Field-induced quantum metal–insulator transition in the
  pyrochlore iridate Nd2Ir2O7}},}\ }\href {\doibase 10.1038/nphys3567}
  {\bibfield  {journal} {\bibinfo  {journal} {Nat. Phys.}\ }\textbf {\bibinfo
  {volume} {12}},\ \bibinfo {pages} {134} (\bibinfo {year} {2015})}\BibitemShut
  {NoStop}%
\bibitem [{\citenamefont {Wakeham}\ \emph {et~al.}(2016)\citenamefont
  {Wakeham}, \citenamefont {Bauer}, \citenamefont {Neupane},\ and\
  \citenamefont {Ronning}}]{Wakeham2016}%
  \BibitemOpen
  \bibfield  {author} {\bibinfo {author} {\bibfnamefont {N.}~\bibnamefont
  {Wakeham}}, \bibinfo {author} {\bibfnamefont {E.~D.}\ \bibnamefont {Bauer}},
  \bibinfo {author} {\bibfnamefont {M.}~\bibnamefont {Neupane}}, \ and\
  \bibinfo {author} {\bibfnamefont {F.}~\bibnamefont {Ronning}},\ }\bibfield
  {title} {\enquote {\bibinfo {title} {{Large magnetoresistance in the
  antiferromagnetic semimetal NdSb}},}\ }\href {\doibase
  10.1103/PhysRevB.93.205152} {\bibfield  {journal} {\bibinfo  {journal} {Phys.
  Rev. B - Condens. Matter Mater. Phys.}\ }\textbf {\bibinfo {volume} {93}},\
  \bibinfo {pages} {205152} (\bibinfo {year} {2016})},\ \Eprint
  {http://arxiv.org/abs/1606.03724} {arXiv:1606.03724} \BibitemShut {NoStop}%
\bibitem [{\citenamefont {Xiao}, \citenamefont {Chang},\ and\ \citenamefont
  {Niu}(2010)}]{Xiao2010b}%
  \BibitemOpen
  \bibfield  {author} {\bibinfo {author} {\bibfnamefont {D.}~\bibnamefont
  {Xiao}}, \bibinfo {author} {\bibfnamefont {M.-C.}\ \bibnamefont {Chang}}, \
  and\ \bibinfo {author} {\bibfnamefont {Q.}~\bibnamefont {Niu}},\ }\bibfield
  {title} {\enquote {\bibinfo {title} {{Berry phase effects on electronic
  properties}},}\ }\href {\doibase 10.1103/RevModPhys.82.1959} {\bibfield
  {journal} {\bibinfo  {journal} {Rev. Mod. Phys.}\ }\textbf {\bibinfo {volume}
  {82}},\ \bibinfo {pages} {1959--2007} (\bibinfo {year} {2010})}\BibitemShut
  {NoStop}%
\bibitem [{\citenamefont {Wan}\ \emph {et~al.}(2011)\citenamefont {Wan},
  \citenamefont {Turner}, \citenamefont {Vishwanath},\ and\ \citenamefont
  {Savrasov}}]{Wan2011}%
  \BibitemOpen
  \bibfield  {author} {\bibinfo {author} {\bibfnamefont {X.}~\bibnamefont
  {Wan}}, \bibinfo {author} {\bibfnamefont {A.~M.}\ \bibnamefont {Turner}},
  \bibinfo {author} {\bibfnamefont {A.}~\bibnamefont {Vishwanath}}, \ and\
  \bibinfo {author} {\bibfnamefont {S.~Y.}\ \bibnamefont {Savrasov}},\
  }\bibfield  {title} {\enquote {\bibinfo {title} {{Topological semimetal and
  Fermi-arc surface states in the electronic structure of pyrochlore
  iridates}},}\ }\href {\doibase 10.1103/PhysRevB.83.205101} {\bibfield
  {journal} {\bibinfo  {journal} {Phys. Rev. B}\ }\textbf {\bibinfo {volume}
  {83}},\ \bibinfo {pages} {205101} (\bibinfo {year} {2011})}\BibitemShut
  {NoStop}%
\bibitem [{\citenamefont {Yang}\ \emph {et~al.}(2016)\citenamefont {Yang},
  \citenamefont {Sun}, \citenamefont {Zhang}, \citenamefont {Shi},
  \citenamefont {Parkin},\ and\ \citenamefont {Yan}}]{Yang2016a}%
  \BibitemOpen
  \bibfield  {author} {\bibinfo {author} {\bibfnamefont {H.}~\bibnamefont
  {Yang}}, \bibinfo {author} {\bibfnamefont {Y.}~\bibnamefont {Sun}}, \bibinfo
  {author} {\bibfnamefont {Y.}~\bibnamefont {Zhang}}, \bibinfo {author}
  {\bibfnamefont {W.-J.}\ \bibnamefont {Shi}}, \bibinfo {author} {\bibfnamefont
  {S.~S.~P.}\ \bibnamefont {Parkin}}, \ and\ \bibinfo {author} {\bibfnamefont
  {B.}~\bibnamefont {Yan}},\ }\bibfield  {title} {\enquote {\bibinfo {title}
  {{Topological Weyl semimetals in the chiral antiferromagnetic materials Mn3Ge
  and Mn3Sn}},}\ }\href {http://arxiv.org/abs/1608.03404} {\bibfield  {journal}
  {\bibinfo  {journal} {arXiv:1608.03404}\ } (\bibinfo {year}
  {2016})}\BibitemShut {NoStop}%
\bibitem [{\citenamefont {Wang}\ \emph
  {et~al.}(2016{\natexlab{c}})\citenamefont {Wang}, \citenamefont {Vergniory},
  \citenamefont {Kushwaha}, \citenamefont {Hirschberger}, \citenamefont
  {Chulkov}, \citenamefont {Ernst}, \citenamefont {Ong}, \citenamefont {Cava},\
  and\ \citenamefont {Bernevig}}]{Wang2016c}%
  \BibitemOpen
  \bibfield  {author} {\bibinfo {author} {\bibfnamefont {Z.}~\bibnamefont
  {Wang}}, \bibinfo {author} {\bibfnamefont {M.~G.}\ \bibnamefont {Vergniory}},
  \bibinfo {author} {\bibfnamefont {S.}~\bibnamefont {Kushwaha}}, \bibinfo
  {author} {\bibfnamefont {M.}~\bibnamefont {Hirschberger}}, \bibinfo {author}
  {\bibfnamefont {E.~V.}\ \bibnamefont {Chulkov}}, \bibinfo {author}
  {\bibfnamefont {A.}~\bibnamefont {Ernst}}, \bibinfo {author} {\bibfnamefont
  {N.~P.}\ \bibnamefont {Ong}}, \bibinfo {author} {\bibfnamefont {R.~J.}\
  \bibnamefont {Cava}}, \ and\ \bibinfo {author} {\bibfnamefont {B.~A.}\
  \bibnamefont {Bernevig}},\ }\bibfield  {title} {\enquote {\bibinfo {title}
  {{Time-Reversal-Breaking Weyl Fermions in Magnetic Heusler Alloys}},}\ }\href
  {\doibase 10.1103/PhysRevLett.117.236401} {\bibfield  {journal} {\bibinfo
  {journal} {Phys. Rev. Lett.}\ }\textbf {\bibinfo {volume} {117}},\ \bibinfo
  {pages} {236401} (\bibinfo {year} {2016}{\natexlab{c}})},\ \Eprint
  {http://arxiv.org/abs/1603.00479} {arXiv:1603.00479} \BibitemShut {NoStop}%
\bibitem [{\citenamefont {Sushkov}\ \emph {et~al.}(2015)\citenamefont
  {Sushkov}, \citenamefont {Hofmann}, \citenamefont {Jenkins}, \citenamefont
  {Ishikawa}, \citenamefont {Nakatsuji}, \citenamefont {{Das Sarma}},\ and\
  \citenamefont {Drew}}]{Sushkov2015}%
  \BibitemOpen
  \bibfield  {author} {\bibinfo {author} {\bibfnamefont {A.~B.}\ \bibnamefont
  {Sushkov}}, \bibinfo {author} {\bibfnamefont {J.~B.}\ \bibnamefont
  {Hofmann}}, \bibinfo {author} {\bibfnamefont {G.~S.}\ \bibnamefont
  {Jenkins}}, \bibinfo {author} {\bibfnamefont {J.}~\bibnamefont {Ishikawa}},
  \bibinfo {author} {\bibfnamefont {S.}~\bibnamefont {Nakatsuji}}, \bibinfo
  {author} {\bibfnamefont {S.}~\bibnamefont {{Das Sarma}}}, \ and\ \bibinfo
  {author} {\bibfnamefont {H.~D.}\ \bibnamefont {Drew}},\ }\bibfield  {title}
  {\enquote {\bibinfo {title} {{Optical evidence for a Weyl semimetal state in
  pyrochlore Eu2Ir2O7}},}\ }\href {\doibase 10.1103/PhysRevB.92.241108}
  {\bibfield  {journal} {\bibinfo  {journal} {Phys. Rev. B - Condens. Matter
  Mater. Phys.}\ }\textbf {\bibinfo {volume} {92}},\ \bibinfo {pages}
  {241108(R)} (\bibinfo {year} {2015})},\ \Eprint
  {http://arxiv.org/abs/1507.01038} {arXiv:1507.01038} \BibitemShut {NoStop}%
\bibitem [{\citenamefont {Yan}\ and\ \citenamefont {Felser}(2017)}]{Yan2016a}%
  \BibitemOpen
  \bibfield  {author} {\bibinfo {author} {\bibfnamefont {B.}~\bibnamefont
  {Yan}}\ and\ \bibinfo {author} {\bibfnamefont {C.}~\bibnamefont {Felser}},\
  }\bibfield  {title} {\enquote {\bibinfo {title} {{Topological Materials: Weyl
  Semimetals}},}\ }\href {\doibase 10.1146/annurev-conmatphys-031016-025458}
  {\bibfield  {journal} {\bibinfo  {journal} {Annu. Rev. Condens. Matter
  Phys.}\ }\textbf {\bibinfo {volume} {8}},\ \bibinfo {pages} {337--354}
  (\bibinfo {year} {2017})},\ \Eprint {http://arxiv.org/abs/1611.04182}
  {arXiv:1611.04182} \BibitemShut {NoStop}%
\bibitem [{\citenamefont {Borisenko}\ \emph {et~al.}(2015)\citenamefont
  {Borisenko}, \citenamefont {Evtushinsky}, \citenamefont {Gibson},
  \citenamefont {Yaresko}, \citenamefont {Kim}, \citenamefont {Ali},
  \citenamefont {Buechner}, \citenamefont {Hoesch},\ and\ \citenamefont
  {Cava}}]{Borisenko2015}%
  \BibitemOpen
  \bibfield  {author} {\bibinfo {author} {\bibfnamefont {S.}~\bibnamefont
  {Borisenko}}, \bibinfo {author} {\bibfnamefont {D.}~\bibnamefont
  {Evtushinsky}}, \bibinfo {author} {\bibfnamefont {Q.}~\bibnamefont {Gibson}},
  \bibinfo {author} {\bibfnamefont {A.}~\bibnamefont {Yaresko}}, \bibinfo
  {author} {\bibfnamefont {T.}~\bibnamefont {Kim}}, \bibinfo {author}
  {\bibfnamefont {M.~N.}\ \bibnamefont {Ali}}, \bibinfo {author} {\bibfnamefont
  {B.}~\bibnamefont {Buechner}}, \bibinfo {author} {\bibfnamefont
  {M.}~\bibnamefont {Hoesch}}, \ and\ \bibinfo {author} {\bibfnamefont {R.~J.}\
  \bibnamefont {Cava}},\ }\bibfield  {title} {\enquote {\bibinfo {title}
  {{Time-Reversal Symmetry Breaking Type-II Weyl State in YbMnBi2}},}\ }\href
  {\doibase 10.1017/CBO9781107415324.004} {\bibfield  {journal} {\bibinfo
  {journal} {arXiv:}\ } (\bibinfo {year} {2015}),\
  10.1017/CBO9781107415324.004},\ \Eprint {http://arxiv.org/abs/1507.04847}
  {arXiv:1507.04847} \BibitemShut {NoStop}%
\bibitem [{\citenamefont {Chinotti}\ \emph {et~al.}(2016)\citenamefont
  {Chinotti}, \citenamefont {Pal}, \citenamefont {Ren}, \citenamefont
  {Petrovic},\ and\ \citenamefont {Degiorgi}}]{Chinotti2016}%
  \BibitemOpen
  \bibfield  {author} {\bibinfo {author} {\bibfnamefont {M.}~\bibnamefont
  {Chinotti}}, \bibinfo {author} {\bibfnamefont {A.}~\bibnamefont {Pal}},
  \bibinfo {author} {\bibfnamefont {W.~J.}\ \bibnamefont {Ren}}, \bibinfo
  {author} {\bibfnamefont {C.}~\bibnamefont {Petrovic}}, \ and\ \bibinfo
  {author} {\bibfnamefont {L.}~\bibnamefont {Degiorgi}},\ }\bibfield  {title}
  {\enquote {\bibinfo {title} {{Electrodynamic response of the type-II Weyl
  semimetal YbMnBi2}},}\ }\href {\doibase 10.1103/PhysRevB.94.245101}
  {\bibfield  {journal} {\bibinfo  {journal} {Phys. Rev. B}\ }\textbf {\bibinfo
  {volume} {94}},\ \bibinfo {pages} {245101} (\bibinfo {year}
  {2016})}\BibitemShut {NoStop}%
\bibitem [{\citenamefont {Wang}\ \emph
  {et~al.}(2016{\natexlab{d}})\citenamefont {Wang}, \citenamefont {Zaliznyak},
  \citenamefont {Ren}, \citenamefont {Wu}, \citenamefont {Graf},\ and\
  \citenamefont {Garlea}}]{Wang2016k}%
  \BibitemOpen
  \bibfield  {author} {\bibinfo {author} {\bibfnamefont {A.}~\bibnamefont
  {Wang}}, \bibinfo {author} {\bibfnamefont {I.}~\bibnamefont {Zaliznyak}},
  \bibinfo {author} {\bibfnamefont {W.}~\bibnamefont {Ren}}, \bibinfo {author}
  {\bibfnamefont {L.}~\bibnamefont {Wu}}, \bibinfo {author} {\bibfnamefont
  {D.}~\bibnamefont {Graf}}, \ and\ \bibinfo {author} {\bibfnamefont {V.~O.}\
  \bibnamefont {Garlea}},\ }\bibfield  {title} {\enquote {\bibinfo {title}
  {{Magnetotransport study of Dirac fermions in YbMnBi2 antiferromagnet}},}\
  }\href {\doibase 10.1103/PhysRevB.94.165161} {\ \textbf {\bibinfo {volume}
  {94}},\ \bibinfo {pages} {165161} (\bibinfo {year}
  {2016}{\natexlab{d}})}\BibitemShut {NoStop}%
\bibitem [{\citenamefont {Chaudhuri}\ \emph {et~al.}()\citenamefont
  {Chaudhuri}, \citenamefont {Cheng}, \citenamefont {Yaresko}, \citenamefont
  {Gibson}, \citenamefont {Cava},\ and\ \citenamefont
  {Armitage}}]{Chaudhuri2017}%
  \BibitemOpen
  \bibfield  {author} {\bibinfo {author} {\bibfnamefont {D.}~\bibnamefont
  {Chaudhuri}}, \bibinfo {author} {\bibfnamefont {B.}~\bibnamefont {Cheng}},
  \bibinfo {author} {\bibfnamefont {A.}~\bibnamefont {Yaresko}}, \bibinfo
  {author} {\bibfnamefont {Q.~D.}\ \bibnamefont {Gibson}}, \bibinfo {author}
  {\bibfnamefont {R.~J.}\ \bibnamefont {Cava}}, \ and\ \bibinfo {author}
  {\bibfnamefont {N.~P.}\ \bibnamefont {Armitage}},\ }\bibfield  {title}
  {\enquote {\bibinfo {title} {{An optical investigation of the strong
  spin-orbital coupled magnetic semimetal YbMnBi2}},}\ }\href@noop {} {\
  }\Eprint {http://arxiv.org/abs/1701.08693v1} {arXiv:1701.08693v1}
  \BibitemShut {NoStop}%
\bibitem [{\citenamefont {Zhang}\ \emph
  {et~al.}(2017{\natexlab{a}})\citenamefont {Zhang}, \citenamefont {Sun},
  \citenamefont {Yang}, \citenamefont {{\v{Z}}elezn{\'{y}}}, \citenamefont
  {Parkin}, \citenamefont {Felser},\ and\ \citenamefont {Yan}}]{Zhang2016d}%
  \BibitemOpen
  \bibfield  {author} {\bibinfo {author} {\bibfnamefont {Y.}~\bibnamefont
  {Zhang}}, \bibinfo {author} {\bibfnamefont {Y.}~\bibnamefont {Sun}}, \bibinfo
  {author} {\bibfnamefont {H.}~\bibnamefont {Yang}}, \bibinfo {author}
  {\bibfnamefont {J.}~\bibnamefont {{\v{Z}}elezn{\'{y}}}}, \bibinfo {author}
  {\bibfnamefont {S.~P.~P.}\ \bibnamefont {Parkin}}, \bibinfo {author}
  {\bibfnamefont {C.}~\bibnamefont {Felser}}, \ and\ \bibinfo {author}
  {\bibfnamefont {B.}~\bibnamefont {Yan}},\ }\bibfield  {title} {\enquote
  {\bibinfo {title} {{Strong anisotropic anomalous Hall effect and spin Hall
  effect in the chiral antiferromagnetic compounds Mn3X (X= Ge, Sn, Ga, Ir, Rh
  and Pt)}},}\ }\href {\doibase 10.1103/PhysRevB.95.075128} {\bibfield
  {journal} {\bibinfo  {journal} {Phys. Rev. B}\ }\textbf {\bibinfo {volume}
  {95}},\ \bibinfo {pages} {075128} (\bibinfo {year}
  {2017}{\natexlab{a}})}\BibitemShut {NoStop}%
\bibitem [{\citenamefont {Kiyohara}, \citenamefont {Tomita},\ and\
  \citenamefont {Nakatsuji}(2016)}]{Kiyohara2015}%
  \BibitemOpen
  \bibfield  {author} {\bibinfo {author} {\bibfnamefont {N.}~\bibnamefont
  {Kiyohara}}, \bibinfo {author} {\bibfnamefont {T.}~\bibnamefont {Tomita}}, \
  and\ \bibinfo {author} {\bibfnamefont {S.}~\bibnamefont {Nakatsuji}},\
  }\bibfield  {title} {\enquote {\bibinfo {title} {{Giant Anomalous Hall Effect
  in the Chiral Antiferromagnet Mn3Ge}},}\ }\href {\doibase
  10.1103/PhysRevApplied.5.064009} {\bibfield  {journal} {\bibinfo  {journal}
  {Phys. Rev. Appl.}\ }\textbf {\bibinfo {volume} {5}},\ \bibinfo {pages}
  {064009} (\bibinfo {year} {2016})},\ \Eprint
  {http://arxiv.org/abs/1511.04619} {arXiv:1511.04619} \BibitemShut {NoStop}%
\bibitem [{\citenamefont {Sinova}\ \emph {et~al.}(2015)\citenamefont {Sinova},
  \citenamefont {Valenzuela}, \citenamefont {Wunderlich}, \citenamefont
  {Back},\ and\ \citenamefont {Jungwirth}}]{Sinova2015}%
  \BibitemOpen
  \bibfield  {author} {\bibinfo {author} {\bibfnamefont {J.}~\bibnamefont
  {Sinova}}, \bibinfo {author} {\bibfnamefont {S.~O.}\ \bibnamefont
  {Valenzuela}}, \bibinfo {author} {\bibfnamefont {J.}~\bibnamefont
  {Wunderlich}}, \bibinfo {author} {\bibfnamefont {C.~H.}\ \bibnamefont
  {Back}}, \ and\ \bibinfo {author} {\bibfnamefont {T.}~\bibnamefont
  {Jungwirth}},\ }\bibfield  {title} {\enquote {\bibinfo {title} {{Spin Hall
  effects}},}\ }\href {\doibase 10.1103/RevModPhys.87.1213} {\bibfield
  {journal} {\bibinfo  {journal} {Rev. Mod. Phys.}\ }\textbf {\bibinfo {volume}
  {87}},\ \bibinfo {pages} {1213--1260} (\bibinfo {year} {2015})},\ \Eprint
  {http://arxiv.org/abs/1411.3249} {arXiv:1411.3249} \BibitemShut {NoStop}%
\bibitem [{\citenamefont {Chen}, \citenamefont {Niu},\ and\ \citenamefont
  {MacDonald}(2014)}]{Chen2014}%
  \BibitemOpen
  \bibfield  {author} {\bibinfo {author} {\bibfnamefont {H.}~\bibnamefont
  {Chen}}, \bibinfo {author} {\bibfnamefont {Q.}~\bibnamefont {Niu}}, \ and\
  \bibinfo {author} {\bibfnamefont {A.~H.}\ \bibnamefont {MacDonald}},\
  }\bibfield  {title} {\enquote {\bibinfo {title} {{Anomalous Hall Effect
  Arising from Noncollinear Antiferromagnetism}},}\ }\href {\doibase
  10.1103/PhysRevLett.112.017205} {\bibfield  {journal} {\bibinfo  {journal}
  {Phys. Rev. Lett.}\ }\textbf {\bibinfo {volume} {112}},\ \bibinfo {pages}
  {017205} (\bibinfo {year} {2014})}\BibitemShut {NoStop}%
\bibitem [{\citenamefont {K{\"{u}}bler}\ and\ \citenamefont
  {Felser}(2014)}]{Kubler2014}%
  \BibitemOpen
  \bibfield  {author} {\bibinfo {author} {\bibfnamefont {J.}~\bibnamefont
  {K{\"{u}}bler}}\ and\ \bibinfo {author} {\bibfnamefont {C.}~\bibnamefont
  {Felser}},\ }\bibfield  {title} {\enquote {\bibinfo {title} {{Non-collinear
  antiferromagnets and the anomalous Hall effect}},}\ }\href {\doibase
  10.1209/0295-5075/108/67001} {\bibfield  {journal} {\bibinfo  {journal}
  {Europhys. Lett.}\ }\textbf {\bibinfo {volume} {108}},\ \bibinfo {pages}
  {67001} (\bibinfo {year} {2014})},\ \Eprint {http://arxiv.org/abs/1410.5985}
  {arXiv:1410.5985} \BibitemShut {NoStop}%
\bibitem [{\citenamefont {Haldane}(1988)}]{Haldane1988}%
  \BibitemOpen
  \bibfield  {author} {\bibinfo {author} {\bibfnamefont {F.~D.~M.}\
  \bibnamefont {Haldane}},\ }\bibfield  {title} {\enquote {\bibinfo {title}
  {{Model for a Quantum Hall Effect without Landau Levels: Condensed-Matter
  Realization of the {\{}`Parity Anomaly'{\}}}},}\ }\href {\doibase
  10.1103/PhysRevLett.61.2015} {\bibfield  {journal} {\bibinfo  {journal}
  {Phys. Rev. Lett.}\ }\textbf {\bibinfo {volume} {61}},\ \bibinfo {pages}
  {2015} (\bibinfo {year} {1988})}\BibitemShut {NoStop}%
\bibitem [{\citenamefont {Shindou}\ and\ \citenamefont
  {Nagaosa}(2001)}]{Shindou2001}%
  \BibitemOpen
  \bibfield  {author} {\bibinfo {author} {\bibfnamefont {R.}~\bibnamefont
  {Shindou}}\ and\ \bibinfo {author} {\bibfnamefont {N.}~\bibnamefont
  {Nagaosa}},\ }\bibfield  {title} {\enquote {\bibinfo {title} {{Orbital
  Ferromagnetism and Anomalous Hall Effect in Antiferromagnets on the Distorted
  fcc Lattice}},}\ }\href {\doibase 10.1103/PhysRevLett.87.116801} {\bibfield
  {journal} {\bibinfo  {journal} {Phys. Rev. Lett.}\ }\textbf {\bibinfo
  {volume} {87}},\ \bibinfo {pages} {116801} (\bibinfo {year}
  {2001})}\BibitemShut {NoStop}%
\bibitem [{\citenamefont {Tomizawa}\ and\ \citenamefont
  {Kontani}(2009)}]{Tomizawa2009}%
  \BibitemOpen
  \bibfield  {author} {\bibinfo {author} {\bibfnamefont {T.}~\bibnamefont
  {Tomizawa}}\ and\ \bibinfo {author} {\bibfnamefont {H.}~\bibnamefont
  {Kontani}},\ }\bibfield  {title} {\enquote {\bibinfo {title} {{Anomalous Hall
  effect in the t2g orbital kagome lattice due to noncollinearity: Significance
  of the orbital Aharonov-Bohm effect}},}\ }\href {\doibase
  10.1103/PhysRevB.80.100401} {\bibfield  {journal} {\bibinfo  {journal} {Phys.
  Rev. B}\ }\textbf {\bibinfo {volume} {80}},\ \bibinfo {pages} {100401(R)}
  (\bibinfo {year} {2009})}\BibitemShut {NoStop}%
\bibitem [{\citenamefont {Tomizawa}\ and\ \citenamefont
  {Kontani}(2010)}]{Tomizawa2010}%
  \BibitemOpen
  \bibfield  {author} {\bibinfo {author} {\bibfnamefont {T.}~\bibnamefont
  {Tomizawa}}\ and\ \bibinfo {author} {\bibfnamefont {H.}~\bibnamefont
  {Kontani}},\ }\bibfield  {title} {\enquote {\bibinfo {title} {{Anomalous Hall
  effect due to noncollinearity in pyrochlore compounds: Role of orbital
  Aharonov-Bohm effect}},}\ }\href {\doibase 10.1103/PhysRevB.82.104412}
  {\bibfield  {journal} {\bibinfo  {journal} {Phys. Rev. B}\ }\textbf {\bibinfo
  {volume} {82}},\ \bibinfo {pages} {104412} (\bibinfo {year}
  {2010})}\BibitemShut {NoStop}%
\bibitem [{\citenamefont {Burkov}\ and\ \citenamefont
  {Balents}(2011)}]{Burkov2011a}%
  \BibitemOpen
  \bibfield  {author} {\bibinfo {author} {\bibfnamefont {A.~A.}\ \bibnamefont
  {Burkov}}\ and\ \bibinfo {author} {\bibfnamefont {L.}~\bibnamefont
  {Balents}},\ }\bibfield  {title} {\enquote {\bibinfo {title} {{Weyl Semimetal
  in a Topological Insulator Multilayer}},}\ }\href {\doibase
  10.1103/PhysRevLett.107.127205} {\bibfield  {journal} {\bibinfo  {journal}
  {Phys. Rev. B}\ }\textbf {\bibinfo {volume} {107}},\ \bibinfo {pages}
  {127205} (\bibinfo {year} {2011})}\BibitemShut {NoStop}%
\bibitem [{\citenamefont {Yang}, \citenamefont {Lu},\ and\ \citenamefont
  {Ran}(2011)}]{Yang2011b}%
  \BibitemOpen
  \bibfield  {author} {\bibinfo {author} {\bibfnamefont {K.~Y.}\ \bibnamefont
  {Yang}}, \bibinfo {author} {\bibfnamefont {Y.~M.}\ \bibnamefont {Lu}}, \ and\
  \bibinfo {author} {\bibfnamefont {Y.}~\bibnamefont {Ran}},\ }\bibfield
  {title} {\enquote {\bibinfo {title} {{Quantum Hall effects in a Weyl
  semimetal: Possible application in pyrochlore iridates}},}\ }\href {\doibase
  10.1103/PhysRevB.84.075129} {\bibfield  {journal} {\bibinfo  {journal} {Phys.
  Rev. B}\ }\textbf {\bibinfo {volume} {84}},\ \bibinfo {pages} {075129}
  (\bibinfo {year} {2011})},\ \Eprint {http://arxiv.org/abs/1105.2353}
  {arXiv:1105.2353} \BibitemShut {NoStop}%
\bibitem [{\citenamefont {Nakatsuji}, \citenamefont {Kiyohara},\ and\
  \citenamefont {Higo}(2015)}]{Nakatsuji2015}%
  \BibitemOpen
  \bibfield  {author} {\bibinfo {author} {\bibfnamefont {S.}~\bibnamefont
  {Nakatsuji}}, \bibinfo {author} {\bibfnamefont {N.}~\bibnamefont {Kiyohara}},
  \ and\ \bibinfo {author} {\bibfnamefont {T.}~\bibnamefont {Higo}},\
  }\bibfield  {title} {\enquote {\bibinfo {title} {{Large anomalous Hall effect
  in a non-collinear antiferromagnet at room temperature}},}\ }\href
  {\doibase 10.1038/nature15723} {\bibfield  {journal} {\bibinfo  {journal}
  {Nature}\ }\textbf {\bibinfo {volume} {527}},\ \bibinfo {pages} {212--216}
  (\bibinfo {year} {2015})}\BibitemShut {NoStop}%
\bibitem [{\citenamefont {Nayak}\ \emph {et~al.}(2016)\citenamefont {Nayak},
  \citenamefont {Fischer}, \citenamefont {Sun}, \citenamefont {Yan},
  \citenamefont {Karel}, \citenamefont {Komarek}, \citenamefont {Shekhar},
  \citenamefont {Kumar}, \citenamefont {Schnelle}, \citenamefont
  {K{\"{u}}bler}, \citenamefont {Felser}, \citenamefont {Parkin}, \citenamefont
  {Ku~bler}, \citenamefont {Felser},\ and\ \citenamefont {Parkin}}]{Nayak2016}%
  \BibitemOpen
  \bibfield  {author} {\bibinfo {author} {\bibfnamefont {A.~K.}\ \bibnamefont
  {Nayak}}, \bibinfo {author} {\bibfnamefont {J.~E.}\ \bibnamefont {Fischer}},
  \bibinfo {author} {\bibfnamefont {Y.}~\bibnamefont {Sun}}, \bibinfo {author}
  {\bibfnamefont {B.}~\bibnamefont {Yan}}, \bibinfo {author} {\bibfnamefont
  {J.}~\bibnamefont {Karel}}, \bibinfo {author} {\bibfnamefont {A.~C.}\
  \bibnamefont {Komarek}}, \bibinfo {author} {\bibfnamefont {C.}~\bibnamefont
  {Shekhar}}, \bibinfo {author} {\bibfnamefont {N.}~\bibnamefont {Kumar}},
  \bibinfo {author} {\bibfnamefont {W.}~\bibnamefont {Schnelle}}, \bibinfo
  {author} {\bibfnamefont {J.}~\bibnamefont {K{\"{u}}bler}}, \bibinfo {author}
  {\bibfnamefont {C.}~\bibnamefont {Felser}}, \bibinfo {author} {\bibfnamefont
  {S.~S.~P.}\ \bibnamefont {Parkin}}, \bibinfo {author} {\bibfnamefont
  {J.}~\bibnamefont {Ku~bler}}, \bibinfo {author} {\bibfnamefont
  {C.}~\bibnamefont {Felser}}, \ and\ \bibinfo {author} {\bibfnamefont
  {S.~S.~P.}\ \bibnamefont {Parkin}},\ }\bibfield  {title} {\enquote {\bibinfo
  {title} {{Large anomalous Hall effect driven by a nonvanishing Berry
  curvature in the noncolinear antiferromagnet Mn3Ge}},}\ }\href {\doibase
  10.1126/sciadv.1501870} {\bibfield  {journal} {\bibinfo  {journal} {Sci.
  Adv.}\ }\textbf {\bibinfo {volume} {2}},\ \bibinfo {pages}
  {e1501870--e1501870} (\bibinfo {year} {2016})},\ \Eprint
  {http://arxiv.org/abs/1511.03128} {arXiv:1511.03128} \BibitemShut {NoStop}%
\bibitem [{\citenamefont {Suzuki}\ \emph {et~al.}(2016)\citenamefont {Suzuki},
  \citenamefont {Chisnell}, \citenamefont {Devarakonda}, \citenamefont {Liu},
  \citenamefont {Feng}, \citenamefont {Xiao}, \citenamefont {Lynn},\ and\
  \citenamefont {Checkelsky}}]{Suzuki2016}%
  \BibitemOpen
  \bibfield  {author} {\bibinfo {author} {\bibfnamefont {T.}~\bibnamefont
  {Suzuki}}, \bibinfo {author} {\bibfnamefont {R.}~\bibnamefont {Chisnell}},
  \bibinfo {author} {\bibfnamefont {A.}~\bibnamefont {Devarakonda}}, \bibinfo
  {author} {\bibfnamefont {Y.-T.}\ \bibnamefont {Liu}}, \bibinfo {author}
  {\bibfnamefont {W.}~\bibnamefont {Feng}}, \bibinfo {author} {\bibfnamefont
  {D.}~\bibnamefont {Xiao}}, \bibinfo {author} {\bibfnamefont {J.~W.}\
  \bibnamefont {Lynn}}, \ and\ \bibinfo {author} {\bibfnamefont {J.~G.}\
  \bibnamefont {Checkelsky}},\ }\bibfield  {title} {\enquote {\bibinfo {title}
  {{Large anomalous Hall effect in a half-Heusler antiferromagnet}},}\ }\href
  {\doibase 10.1038/nphys3831} {\bibfield  {journal} {\bibinfo  {journal} {Nat.
  Phys.}\ }\textbf {\bibinfo {volume} {12}},\ \bibinfo {pages} {1119} (\bibinfo
  {year} {2016})}\BibitemShut {NoStop}%
\bibitem [{\citenamefont {Zhou}, \citenamefont {Sun},\ and\ \citenamefont
  {Sun}(2016)}]{Zhou2016a}%
  \BibitemOpen
  \bibfield  {author} {\bibinfo {author} {\bibfnamefont {P.}~\bibnamefont
  {Zhou}}, \bibinfo {author} {\bibfnamefont {C.~Q.}\ \bibnamefont {Sun}}, \
  and\ \bibinfo {author} {\bibfnamefont {L.~Z.}\ \bibnamefont {Sun}},\
  }\bibfield  {title} {\enquote {\bibinfo {title} {{Two Dimensional
  Antiferromagnetic Chern Insulator: NiRuCl6}},}\ }\href {\doibase
  10.1021/acs.nanolett.6b02701} {\bibfield  {journal} {\bibinfo  {journal}
  {Nano Lett.}\ }\textbf {\bibinfo {volume} {16}},\ \bibinfo {pages}
  {6325--6330} (\bibinfo {year} {2016})},\ \Eprint
  {http://arxiv.org/abs/1601.07705} {arXiv:1601.07705} \BibitemShut {NoStop}%
\bibitem [{\citenamefont {Dong}\ \emph {et~al.}(2016)\citenamefont {Dong},
  \citenamefont {Kanungo}, \citenamefont {Yan},\ and\ \citenamefont
  {Liu}}]{Dong2016}%
  \BibitemOpen
  \bibfield  {author} {\bibinfo {author} {\bibfnamefont {X.-Y.}\ \bibnamefont
  {Dong}}, \bibinfo {author} {\bibfnamefont {S.}~\bibnamefont {Kanungo}},
  \bibinfo {author} {\bibfnamefont {B.}~\bibnamefont {Yan}}, \ and\ \bibinfo
  {author} {\bibfnamefont {C.-X.}\ \bibnamefont {Liu}},\ }\bibfield  {title}
  {\enquote {\bibinfo {title} {{Time-reversal-breaking topological phases in
  antiferromagnetic Sr2FeOsO6 films}},}\ }\href {\doibase
  10.1103/PhysRevB.94.245135} {\bibfield  {journal} {\bibinfo  {journal} {Phys.
  Rev. B}\ }\textbf {\bibinfo {volume} {94}},\ \bibinfo {pages} {245135}
  (\bibinfo {year} {2016})}\BibitemShut {NoStop}%
\bibitem [{\citenamefont {Hanke}\ \emph
  {et~al.}(2017{\natexlab{a}})\citenamefont {Hanke}, \citenamefont {Freimuth},
  \citenamefont {Niu}, \citenamefont {Bl{\"{u}}gel},\ and\ \citenamefont
  {Mokrousov}}]{Hanke2017a}%
  \BibitemOpen
  \bibfield  {author} {\bibinfo {author} {\bibfnamefont {J.-P.}\ \bibnamefont
  {Hanke}}, \bibinfo {author} {\bibfnamefont {F.}~\bibnamefont {Freimuth}},
  \bibinfo {author} {\bibfnamefont {C.}~\bibnamefont {Niu}}, \bibinfo {author}
  {\bibfnamefont {S.}~\bibnamefont {Bl{\"{u}}gel}}, \ and\ \bibinfo {author}
  {\bibfnamefont {Y.}~\bibnamefont {Mokrousov}},\ }\bibfield  {title} {\enquote
  {\bibinfo {title} {{Mixed Weyl semimetals and dissipationless magnetization
  control in insulators}},}\ }\href {http://arxiv.org/abs/1701.08050}
  {\bibfield  {journal} {\bibinfo  {journal} {arxiv.org/1701.08050}\ }
  (\bibinfo {year} {2017}{\natexlab{a}})},\ \Eprint
  {http://arxiv.org/abs/1701.08050} {arXiv:1701.08050} \BibitemShut {NoStop}%
\bibitem [{\citenamefont {Li}\ \emph {et~al.}(2016)\citenamefont {Li},
  \citenamefont {Li}, \citenamefont {Kim}, \citenamefont {Balents},
  \citenamefont {Yu},\ and\ \citenamefont {Chen}}]{Li2016a}%
  \BibitemOpen
  \bibfield  {author} {\bibinfo {author} {\bibfnamefont {F.-Y.}\ \bibnamefont
  {Li}}, \bibinfo {author} {\bibfnamefont {Y.-D.}\ \bibnamefont {Li}}, \bibinfo
  {author} {\bibfnamefont {Y.~B.}\ \bibnamefont {Kim}}, \bibinfo {author}
  {\bibfnamefont {L.}~\bibnamefont {Balents}}, \bibinfo {author} {\bibfnamefont
  {Y.}~\bibnamefont {Yu}}, \ and\ \bibinfo {author} {\bibfnamefont
  {G.}~\bibnamefont {Chen}},\ }\bibfield  {title} {\enquote {\bibinfo {title}
  {{Weyl magnons in breathing pyrochlore antiferromagnets}},}\ }\href {\doibase
  10.1038/ncomms12691} {\bibfield  {journal} {\bibinfo  {journal} {Nat.
  Commun.}\ }\textbf {\bibinfo {volume} {7}},\ \bibinfo {pages} {12691}
  (\bibinfo {year} {2016})},\ \Eprint {http://arxiv.org/abs/1602.04288}
  {arXiv:1602.04288} \BibitemShut {NoStop}%
\bibitem [{\citenamefont {Kanazawa}\ \emph {et~al.}(2011)\citenamefont
  {Kanazawa}, \citenamefont {Onose}, \citenamefont {Arima}, \citenamefont
  {Okuyama}, \citenamefont {Ohoyama}, \citenamefont {Wakimoto}, \citenamefont
  {Kakurai}, \citenamefont {Ishiwata},\ and\ \citenamefont
  {Tokura}}]{Kanazawa2011}%
  \BibitemOpen
  \bibfield  {author} {\bibinfo {author} {\bibfnamefont {N.}~\bibnamefont
  {Kanazawa}}, \bibinfo {author} {\bibfnamefont {Y.}~\bibnamefont {Onose}},
  \bibinfo {author} {\bibfnamefont {T.}~\bibnamefont {Arima}}, \bibinfo
  {author} {\bibfnamefont {D.}~\bibnamefont {Okuyama}}, \bibinfo {author}
  {\bibfnamefont {K.}~\bibnamefont {Ohoyama}}, \bibinfo {author} {\bibfnamefont
  {S.}~\bibnamefont {Wakimoto}}, \bibinfo {author} {\bibfnamefont
  {K.}~\bibnamefont {Kakurai}}, \bibinfo {author} {\bibfnamefont
  {S.}~\bibnamefont {Ishiwata}}, \ and\ \bibinfo {author} {\bibfnamefont
  {Y.}~\bibnamefont {Tokura}},\ }\bibfield  {title} {\enquote {\bibinfo {title}
  {{Large Topological Hall Effect in a Short-Period Helimagnet MnGe}},}\ }\href
  {\doibase 10.1103/PhysRevLett.106.156603} {\bibfield  {journal} {\bibinfo
  {journal} {Phys. Rev. Lett.}\ }\textbf {\bibinfo {volume} {106}},\ \bibinfo
  {pages} {156603} (\bibinfo {year} {2011})}\BibitemShut {NoStop}%
\bibitem [{\citenamefont {Hoffmann}\ \emph {et~al.}(2015)\citenamefont
  {Hoffmann}, \citenamefont {Weischenberg}, \citenamefont {Dup{\'{e}}},
  \citenamefont {Freimuth}, \citenamefont {Ferriani}, \citenamefont
  {Mokrousov},\ and\ \citenamefont {Heinze}}]{Hoffmann2015}%
  \BibitemOpen
  \bibfield  {author} {\bibinfo {author} {\bibfnamefont {M.}~\bibnamefont
  {Hoffmann}}, \bibinfo {author} {\bibfnamefont {J.}~\bibnamefont
  {Weischenberg}}, \bibinfo {author} {\bibfnamefont {B.}~\bibnamefont
  {Dup{\'{e}}}}, \bibinfo {author} {\bibfnamefont {F.}~\bibnamefont
  {Freimuth}}, \bibinfo {author} {\bibfnamefont {P.}~\bibnamefont {Ferriani}},
  \bibinfo {author} {\bibfnamefont {Y.}~\bibnamefont {Mokrousov}}, \ and\
  \bibinfo {author} {\bibfnamefont {S.}~\bibnamefont {Heinze}},\ }\bibfield
  {title} {\enquote {\bibinfo {title} {{Topological orbital magnetization and
  emergent Hall effect of an atomic-scale spin lattice at a surface}},}\ }\href
  {\doibase 10.1103/PhysRevB.92.020401} {\bibfield  {journal} {\bibinfo
  {journal} {Phys. Rev. B}\ }\textbf {\bibinfo {volume} {92}},\ \bibinfo
  {pages} {020401} (\bibinfo {year} {2015})},\ \Eprint
  {http://arxiv.org/abs/1503.01885v2} {arXiv:1503.01885v2} \BibitemShut
  {NoStop}%
\bibitem [{\citenamefont {Machida}\ \emph {et~al.}(2010)\citenamefont
  {Machida}, \citenamefont {Nakatsuji}, \citenamefont {Onoda}, \citenamefont
  {Tayama},\ and\ \citenamefont {Sakakibara}}]{Machida2010}%
  \BibitemOpen
  \bibfield  {author} {\bibinfo {author} {\bibfnamefont {Y.}~\bibnamefont
  {Machida}}, \bibinfo {author} {\bibfnamefont {S.}~\bibnamefont {Nakatsuji}},
  \bibinfo {author} {\bibfnamefont {S.}~\bibnamefont {Onoda}}, \bibinfo
  {author} {\bibfnamefont {T.}~\bibnamefont {Tayama}}, \ and\ \bibinfo {author}
  {\bibfnamefont {T.}~\bibnamefont {Sakakibara}},\ }\bibfield  {title}
  {\enquote {\bibinfo {title} {{Time-reversal symmetry breaking and spontaneous
  Hall effect without magnetic dipole order.}}}\ }\href {\doibase
  10.1038/nature08680} {\bibfield  {journal} {\bibinfo  {journal} {Nature}\
  }\textbf {\bibinfo {volume} {463}},\ \bibinfo {pages} {210--213} (\bibinfo
  {year} {2010})}\BibitemShut {NoStop}%
\bibitem [{\citenamefont {S{\"{u}}rgers}\ \emph {et~al.}(2014)\citenamefont
  {S{\"{u}}rgers}, \citenamefont {Fischer}, \citenamefont {Winkel},\ and\
  \citenamefont {L{\"{o}}hneysen}}]{Surgers2014}%
  \BibitemOpen
  \bibfield  {author} {\bibinfo {author} {\bibfnamefont {C.}~\bibnamefont
  {S{\"{u}}rgers}}, \bibinfo {author} {\bibfnamefont {G.}~\bibnamefont
  {Fischer}}, \bibinfo {author} {\bibfnamefont {P.}~\bibnamefont {Winkel}}, \
  and\ \bibinfo {author} {\bibfnamefont {H.~V.}\ \bibnamefont
  {L{\"{o}}hneysen}},\ }\bibfield  {title} {\enquote {\bibinfo {title} {{Large
  topological Hall effect in the non-collinear phase of an antiferromagnet.}}}\
  }\href {\doibase 10.1038/ncomms4400} {\bibfield  {journal} {\bibinfo
  {journal} {Nat. Commun.}\ }\textbf {\bibinfo {volume} {5}},\ \bibinfo {pages}
  {3400} (\bibinfo {year} {2014})}\BibitemShut {NoStop}%
\bibitem [{\citenamefont {S{\"{u}}rgers}\ \emph {et~al.}(2016)\citenamefont
  {S{\"{u}}rgers}, \citenamefont {Kittler}, \citenamefont {Wolf},\ and\
  \citenamefont {L{\"{o}}hneysen}}]{Surgers2016}%
  \BibitemOpen
  \bibfield  {author} {\bibinfo {author} {\bibfnamefont {C.}~\bibnamefont
  {S{\"{u}}rgers}}, \bibinfo {author} {\bibfnamefont {W.}~\bibnamefont
  {Kittler}}, \bibinfo {author} {\bibfnamefont {T.}~\bibnamefont {Wolf}}, \
  and\ \bibinfo {author} {\bibfnamefont {H.~v.}\ \bibnamefont
  {L{\"{o}}hneysen}},\ }\bibfield  {title} {\enquote {\bibinfo {title}
  {{Anomalous Hall effect in the noncollinear antiferromagnet Mn5Si3}},}\
  }\href {\doibase 10.1063/1.4943759} {\bibfield  {journal} {\bibinfo
  {journal} {AIP Adv.}\ }\textbf {\bibinfo {volume} {6}},\ \bibinfo {pages}
  {055604} (\bibinfo {year} {2016})},\ \Eprint
  {http://arxiv.org/abs/1601.01840} {arXiv:1601.01840} \BibitemShut {NoStop}%
\bibitem [{\citenamefont {Zhou}\ \emph {et~al.}(2016)\citenamefont {Zhou},
  \citenamefont {Liang}, \citenamefont {Weng}, \citenamefont {Chen},
  \citenamefont {Yao}, \citenamefont {Chen}, \citenamefont {Dong},\ and\
  \citenamefont {Guo}}]{Zhou2016}%
  \BibitemOpen
  \bibfield  {author} {\bibinfo {author} {\bibfnamefont {J.}~\bibnamefont
  {Zhou}}, \bibinfo {author} {\bibfnamefont {Q.~F.}\ \bibnamefont {Liang}},
  \bibinfo {author} {\bibfnamefont {H.}~\bibnamefont {Weng}}, \bibinfo {author}
  {\bibfnamefont {Y.~B.}\ \bibnamefont {Chen}}, \bibinfo {author}
  {\bibfnamefont {S.~H.}\ \bibnamefont {Yao}}, \bibinfo {author} {\bibfnamefont
  {Y.~F.}\ \bibnamefont {Chen}}, \bibinfo {author} {\bibfnamefont
  {J.}~\bibnamefont {Dong}}, \ and\ \bibinfo {author} {\bibfnamefont {G.~Y.}\
  \bibnamefont {Guo}},\ }\bibfield  {title} {\enquote {\bibinfo {title}
  {{Predicted Quantum Topological Hall Effect and Noncoplanar
  Antiferromagnetism in K0.5RhO2}},}\ }\href {\doibase
  10.1103/PhysRevLett.116.256601} {\bibfield  {journal} {\bibinfo  {journal}
  {Phys. Rev. Lett.}\ }\textbf {\bibinfo {volume} {116}},\ \bibinfo {pages}
  {256601} (\bibinfo {year} {2016})},\ \Eprint
  {http://arxiv.org/abs/1602.08553} {arXiv:1602.08553} \BibitemShut {NoStop}%
\bibitem [{\citenamefont {Sun}\ \emph {et~al.}(2016)\citenamefont {Sun},
  \citenamefont {Zhang}, \citenamefont {Felser},\ and\ \citenamefont
  {Yan}}]{Sun2016a}%
  \BibitemOpen
  \bibfield  {author} {\bibinfo {author} {\bibfnamefont {Y.}~\bibnamefont
  {Sun}}, \bibinfo {author} {\bibfnamefont {Y.}~\bibnamefont {Zhang}}, \bibinfo
  {author} {\bibfnamefont {C.}~\bibnamefont {Felser}}, \ and\ \bibinfo {author}
  {\bibfnamefont {B.}~\bibnamefont {Yan}},\ }\bibfield  {title} {\enquote
  {\bibinfo {title} {{Strong Intrinsic Spin Hall Effect in the TaAs Family of
  Weyl Semimetals}},}\ }\href {\doibase 10.1103/PhysRevLett.117.146403}
  {\bibfield  {journal} {\bibinfo  {journal} {Phys. Rev. Lett.}\ }\textbf
  {\bibinfo {volume} {117}},\ \bibinfo {pages} {146403} (\bibinfo {year}
  {2016})},\ \Eprint {http://arxiv.org/abs/1604.07167} {arXiv:1604.07167}
  \BibitemShut {NoStop}%
\bibitem [{\citenamefont {Yin}\ \emph {et~al.}(2015)\citenamefont {Yin},
  \citenamefont {Liu}, \citenamefont {Barlas}, \citenamefont {Zang},\ and\
  \citenamefont {Lake}}]{Yin2015}%
  \BibitemOpen
  \bibfield  {author} {\bibinfo {author} {\bibfnamefont {G.}~\bibnamefont
  {Yin}}, \bibinfo {author} {\bibfnamefont {Y.}~\bibnamefont {Liu}}, \bibinfo
  {author} {\bibfnamefont {Y.}~\bibnamefont {Barlas}}, \bibinfo {author}
  {\bibfnamefont {J.}~\bibnamefont {Zang}}, \ and\ \bibinfo {author}
  {\bibfnamefont {R.~K.}\ \bibnamefont {Lake}},\ }\bibfield  {title} {\enquote
  {\bibinfo {title} {{Topological spin Hall effect resulting from magnetic
  skyrmions}},}\ }\href {\doibase 10.1103/PhysRevB.92.024411} {\bibfield
  {journal} {\bibinfo  {journal} {Phys. Rev. B}\ }\textbf {\bibinfo {volume}
  {92}},\ \bibinfo {pages} {024411} (\bibinfo {year} {2015})},\ \Eprint
  {http://arxiv.org/abs/1503.00242} {arXiv:1503.00242} \BibitemShut {NoStop}%
\bibitem [{\citenamefont {Zhang}\ \emph
  {et~al.}(2017{\natexlab{b}})\citenamefont {Zhang}, \citenamefont {Zelezny},
  \citenamefont {Sun}, \citenamefont {van~den Brink},\ and\ \citenamefont
  {Yan}}]{Zhang2017b}%
  \BibitemOpen
  \bibfield  {author} {\bibinfo {author} {\bibfnamefont {Y.}~\bibnamefont
  {Zhang}}, \bibinfo {author} {\bibfnamefont {J.}~\bibnamefont {Zelezny}},
  \bibinfo {author} {\bibfnamefont {Y.}~\bibnamefont {Sun}}, \bibinfo {author}
  {\bibfnamefont {J.}~\bibnamefont {van~den Brink}}, \ and\ \bibinfo {author}
  {\bibfnamefont {B.}~\bibnamefont {Yan}},\ }\bibfield  {title} {\enquote
  {\bibinfo {title} {{Spin Hall effect emerging from a chiral magnetic lattice
  without spin-orbit coupling}},}\ }\href {http://arxiv.org/abs/1704.03917}
  {\bibfield  {journal} {\bibinfo  {journal} {Arxiv Prepr.}\ } (\bibinfo {year}
  {2017}{\natexlab{b}})},\ \Eprint {http://arxiv.org/abs/1704.03917}
  {arXiv:1704.03917} \BibitemShut {NoStop}%
\bibitem [{\citenamefont {Hanke}\ \emph {et~al.}(2016)\citenamefont {Hanke},
  \citenamefont {Freimuth}, \citenamefont {Nandy}, \citenamefont {Zhang},
  \citenamefont {Bl$\backslash$"{\{}u{\}}gel},\ and\ \citenamefont
  {Mokrousov}}]{Hanke2016}%
  \BibitemOpen
  \bibfield  {author} {\bibinfo {author} {\bibfnamefont {J.~P.}\ \bibnamefont
  {Hanke}}, \bibinfo {author} {\bibfnamefont {F.}~\bibnamefont {Freimuth}},
  \bibinfo {author} {\bibfnamefont {A.~K.}\ \bibnamefont {Nandy}}, \bibinfo
  {author} {\bibfnamefont {H.}~\bibnamefont {Zhang}}, \bibinfo {author}
  {\bibfnamefont {S.}~\bibnamefont {Bl$\backslash$"{\{}u{\}}gel}}, \ and\
  \bibinfo {author} {\bibfnamefont {Y.}~\bibnamefont {Mokrousov}},\ }\bibfield
  {title} {\enquote {\bibinfo {title} {{Role of Berry phase theory for
  describing orbital magnetism: From magnetic heterostructures to topological
  orbital ferromagnets}},}\ }\href {\doibase 10.1103/PhysRevB.94.121114}
  {\bibfield  {journal} {\bibinfo  {journal} {Phys. Rev. B}\ }\textbf {\bibinfo
  {volume} {94}},\ \bibinfo {pages} {121114(R)} (\bibinfo {year} {2016})},\
  \Eprint {http://arxiv.org/abs/1603.07683} {arXiv:1603.07683} \BibitemShut
  {NoStop}%
\bibitem [{\citenamefont {Hanke}\ \emph
  {et~al.}(2017{\natexlab{b}})\citenamefont {Hanke}, \citenamefont {Freimuth},
  \citenamefont {Bl{\"{u}}gel},\ and\ \citenamefont {Mokrousov}}]{Hanke2017b}%
  \BibitemOpen
  \bibfield  {author} {\bibinfo {author} {\bibfnamefont {J.-P.}\ \bibnamefont
  {Hanke}}, \bibinfo {author} {\bibfnamefont {F.}~\bibnamefont {Freimuth}},
  \bibinfo {author} {\bibfnamefont {S.}~\bibnamefont {Bl{\"{u}}gel}}, \ and\
  \bibinfo {author} {\bibfnamefont {Y.}~\bibnamefont {Mokrousov}},\ }\bibfield
  {title} {\enquote {\bibinfo {title} {{Prototypical topological orbital
  ferromagnet $\gamma$-FeMn}},}\ }\href {\doibase 10.1038/srep41078} {\bibfield
   {journal} {\bibinfo  {journal} {Sci. Rep.}\ }\textbf {\bibinfo {volume}
  {7}},\ \bibinfo {pages} {41078} (\bibinfo {year}
  {2017}{\natexlab{b}})}\BibitemShut {NoStop}%
\bibitem [{\citenamefont {Buhl}\ \emph {et~al.}(2017)\citenamefont {Buhl},
  \citenamefont {Freimuth}, \citenamefont {Bl{\"{u}}gel},\ and\ \citenamefont
  {Mokrousov}}]{Buhl2017}%
  \BibitemOpen
  \bibfield  {author} {\bibinfo {author} {\bibfnamefont {P.~M.}\ \bibnamefont
  {Buhl}}, \bibinfo {author} {\bibfnamefont {F.}~\bibnamefont {Freimuth}},
  \bibinfo {author} {\bibfnamefont {S.}~\bibnamefont {Bl{\"{u}}gel}}, \ and\
  \bibinfo {author} {\bibfnamefont {Y.}~\bibnamefont {Mokrousov}},\ }\bibfield
  {title} {\enquote {\bibinfo {title} {{Topological spin Hall effect in
  antiferromagnetic skyrmions}},}\ }\href {\doibase 10.1002/pssr.201700007}
  {\bibfield  {journal} {\bibinfo  {journal} {Phys. status solidi - Rapid Res.
  Lett.}\ }\textbf {\bibinfo {volume} {11}},\ \bibinfo {pages} {1700007}
  (\bibinfo {year} {2017})}\BibitemShut {NoStop}%
\bibitem [{\citenamefont {Jin}\ \emph {et~al.}(2016)\citenamefont {Jin},
  \citenamefont {Song}, \citenamefont {Wang},\ and\ \citenamefont
  {Liu}}]{Jin2016}%
  \BibitemOpen
  \bibfield  {author} {\bibinfo {author} {\bibfnamefont {C.}~\bibnamefont
  {Jin}}, \bibinfo {author} {\bibfnamefont {C.}~\bibnamefont {Song}}, \bibinfo
  {author} {\bibfnamefont {J.}~\bibnamefont {Wang}}, \ and\ \bibinfo {author}
  {\bibfnamefont {Q.}~\bibnamefont {Liu}},\ }\bibfield  {title} {\enquote
  {\bibinfo {title} {{Dynamics of antiferromagnetic skyrmion driven by the spin
  Hall effect}},}\ }\href {\doibase 10.1063/1.4967006} {\bibfield  {journal}
  {\bibinfo  {journal} {Appl. Phys. Lett.}\ }\textbf {\bibinfo {volume}
  {109}},\ \bibinfo {pages} {182404} (\bibinfo {year} {2016})}\BibitemShut
  {NoStop}%
\bibitem [{\citenamefont {Finocchio}\ \emph {et~al.}(2016)\citenamefont
  {Finocchio}, \citenamefont {B{\"{u}}ttner}, \citenamefont {Tomasello},
  \citenamefont {Carpentieri},\ and\ \citenamefont
  {Kl{\"{a}}ui}}]{Finocchio2016}%
  \BibitemOpen
  \bibfield  {author} {\bibinfo {author} {\bibfnamefont {G.}~\bibnamefont
  {Finocchio}}, \bibinfo {author} {\bibfnamefont {F.}~\bibnamefont
  {B{\"{u}}ttner}}, \bibinfo {author} {\bibfnamefont {R.}~\bibnamefont
  {Tomasello}}, \bibinfo {author} {\bibfnamefont {M.}~\bibnamefont
  {Carpentieri}}, \ and\ \bibinfo {author} {\bibfnamefont {M.}~\bibnamefont
  {Kl{\"{a}}ui}},\ }\bibfield  {title} {\enquote {\bibinfo {title} {{Magnetic
  skyrmions: from fundamental to applications}},}\ }\href {\doibase
  10.1088/0022-3727/49/42/423001} {\bibfield  {journal} {\bibinfo  {journal}
  {J. Phys. D. Appl. Phys.}\ }\textbf {\bibinfo {volume} {49}},\ \bibinfo
  {pages} {423001} (\bibinfo {year} {2016})}\BibitemShut {NoStop}%
\bibitem [{\citenamefont {Liu}\ and\ \citenamefont {Ian}(2016)}]{Liu2016a}%
  \BibitemOpen
  \bibfield  {author} {\bibinfo {author} {\bibfnamefont {Z.}~\bibnamefont
  {Liu}}\ and\ \bibinfo {author} {\bibfnamefont {H.}~\bibnamefont {Ian}},\
  }\bibfield  {title} {\enquote {\bibinfo {title} {{Numerical studies on
  antiferromagnetic skyrmions in nanodisks by means of a new quantum simulation
  approach}},}\ }\href {\doibase 10.1016/j.cplett.2016.02.054} {\bibfield
  {journal} {\bibinfo  {journal} {Chem. Phys. Lett.}\ }\textbf {\bibinfo
  {volume} {649}},\ \bibinfo {pages} {135--140} (\bibinfo {year} {2016})},\
  \Eprint {http://arxiv.org/abs/1601.05170v1} {arXiv:1601.05170v1} \BibitemShut
  {NoStop}%
\bibitem [{\citenamefont {Rohart}, \citenamefont {Miltat},\ and\ \citenamefont
  {Thiaville}(2016)}]{Rohart2016}%
  \BibitemOpen
  \bibfield  {author} {\bibinfo {author} {\bibfnamefont {S.}~\bibnamefont
  {Rohart}}, \bibinfo {author} {\bibfnamefont {J.}~\bibnamefont {Miltat}}, \
  and\ \bibinfo {author} {\bibfnamefont {A.}~\bibnamefont {Thiaville}},\
  }\bibfield  {title} {\enquote {\bibinfo {title} {{Path to collapse for an
  isolated N{\'{e}}el skyrmion}},}\ }\href {\doibase
  10.1103/PhysRevB.93.214412} {\bibfield  {journal} {\bibinfo  {journal} {Phys.
  Rev. B}\ }\textbf {\bibinfo {volume} {93}},\ \bibinfo {pages} {214412}
  (\bibinfo {year} {2016})}\BibitemShut {NoStop}%
\bibitem [{\citenamefont {Bogdanov}\ \emph {et~al.}(2002)\citenamefont
  {Bogdanov}, \citenamefont {Roessler}, \citenamefont {Wolf},\ and\
  \citenamefont {Muller}}]{Bogdanov2002}%
  \BibitemOpen
  \bibfield  {author} {\bibinfo {author} {\bibfnamefont {A.~N.}\ \bibnamefont
  {Bogdanov}}, \bibinfo {author} {\bibfnamefont {U.~K.}\ \bibnamefont
  {Roessler}}, \bibinfo {author} {\bibfnamefont {M.}~\bibnamefont {Wolf}}, \
  and\ \bibinfo {author} {\bibfnamefont {K.~H.}\ \bibnamefont {Muller}},\
  }\bibfield  {title} {\enquote {\bibinfo {title} {{Magnetic structures and
  reorientation transitions in noncentrosymmetric uniaxial
  antiferromagnets}},}\ }\href {\doibase 10.1103/PhysRevB.66.214410} {\bibfield
   {journal} {\bibinfo  {journal} {Phys. Rev. B}\ }\textbf {\bibinfo {volume}
  {66}},\ \bibinfo {pages} {214410} (\bibinfo {year} {2002})},\ \Eprint
  {http://arxiv.org/abs/0206291} {arXiv:0206291 [cond-mat]} \BibitemShut
  {NoStop}%
\bibitem [{\citenamefont {Morinari}(2010)}]{Morinari2010}%
  \BibitemOpen
  \bibfield  {author} {\bibinfo {author} {\bibfnamefont {T.}~\bibnamefont
  {Morinari}},\ }\bibfield  {title} {\enquote {\bibinfo {title} {{Half-Skyrmion
  Theory for High-Temperature Superconductivity}},}\ }in\ \href {\doibase
  10.1142/9789814280709_0013} {\emph {\bibinfo {booktitle} {The Multifaceted
  Skyrmion}}}\ (\bibinfo  {publisher} {World Scientific},\ \bibinfo {year}
  {2010})\ pp.\ \bibinfo {pages} {311--331}\BibitemShut {NoStop}%
\bibitem [{\citenamefont {Zhang}, \citenamefont {Zhou},\ and\ \citenamefont
  {Ezawa}(2016)}]{Zhang2015b}%
  \BibitemOpen
  \bibfield  {author} {\bibinfo {author} {\bibfnamefont {X.}~\bibnamefont
  {Zhang}}, \bibinfo {author} {\bibfnamefont {Y.}~\bibnamefont {Zhou}}, \ and\
  \bibinfo {author} {\bibfnamefont {M.}~\bibnamefont {Ezawa}},\ }\bibfield
  {title} {\enquote {\bibinfo {title} {{Antiferromagnetic Skyrmion: Stability,
  Creation and Manipulation}},}\ }\href {\doibase 10.1038/srep24795} {\bibfield
   {journal} {\bibinfo  {journal} {Sci. Rep.}\ }\textbf {\bibinfo {volume}
  {6}},\ \bibinfo {pages} {24795} (\bibinfo {year} {2016})},\ \Eprint
  {http://arxiv.org/abs/1504.01198} {arXiv:1504.01198} \BibitemShut {NoStop}%
\bibitem [{\citenamefont {Barker}\ and\ \citenamefont
  {Tretiakov}(2016)}]{Barker2016}%
  \BibitemOpen
  \bibfield  {author} {\bibinfo {author} {\bibfnamefont {J.}~\bibnamefont
  {Barker}}\ and\ \bibinfo {author} {\bibfnamefont {O.~A.}\ \bibnamefont
  {Tretiakov}},\ }\bibfield  {title} {\enquote {\bibinfo {title} {{Static and
  Dynamical Properties of Antiferromagnetic Skyrmions in the Presence of
  Applied Current and Temperature}},}\ }\href {\doibase
  10.1103/PhysRevLett.116.147203} {\bibfield  {journal} {\bibinfo  {journal}
  {Phys. Rev. Lett.}\ }\textbf {\bibinfo {volume} {116}},\ \bibinfo {pages}
  {147203} (\bibinfo {year} {2016})}\BibitemShut {NoStop}%
\bibitem [{\citenamefont {Velkov}\ \emph {et~al.}(2016)\citenamefont {Velkov},
  \citenamefont {Gomonay}, \citenamefont {Beens}, \citenamefont {Schwiete},
  \citenamefont {Brataas}, \citenamefont {Sinova},\ and\ \citenamefont
  {Duine}}]{Velkov2016}%
  \BibitemOpen
  \bibfield  {author} {\bibinfo {author} {\bibfnamefont {H.}~\bibnamefont
  {Velkov}}, \bibinfo {author} {\bibfnamefont {O.}~\bibnamefont {Gomonay}},
  \bibinfo {author} {\bibfnamefont {M.}~\bibnamefont {Beens}}, \bibinfo
  {author} {\bibfnamefont {G.}~\bibnamefont {Schwiete}}, \bibinfo {author}
  {\bibfnamefont {A.}~\bibnamefont {Brataas}}, \bibinfo {author} {\bibfnamefont
  {J.}~\bibnamefont {Sinova}}, \ and\ \bibinfo {author} {\bibfnamefont {R.~A.}\
  \bibnamefont {Duine}},\ }\bibfield  {title} {\enquote {\bibinfo {title}
  {{Phenomenology of current-induced skyrmion motion in antiferromagnets}},}\
  }\href {\doibase 10.1088/1367-2630/18/7/075016} {\bibfield  {journal}
  {\bibinfo  {journal} {New J. Phys.}\ }\textbf {\bibinfo {volume} {18}},\
  \bibinfo {pages} {075016} (\bibinfo {year} {2016})},\ \Eprint
  {http://arxiv.org/abs/1604.05712} {arXiv:1604.05712} \BibitemShut {NoStop}%
\bibitem [{\citenamefont {Zhang}, \citenamefont {Zhou},\ and\ \citenamefont
  {Ezawa}(2015)}]{Zhang2016h}%
  \BibitemOpen
  \bibfield  {author} {\bibinfo {author} {\bibfnamefont {X.}~\bibnamefont
  {Zhang}}, \bibinfo {author} {\bibfnamefont {Y.}~\bibnamefont {Zhou}}, \ and\
  \bibinfo {author} {\bibfnamefont {M.}~\bibnamefont {Ezawa}},\ }\bibfield
  {title} {\enquote {\bibinfo {title} {{Magnetic bilayer-skyrmions without
  skyrmion Hall effect}},}\ }\href {\doibase 10.1038/ncomms10293} {\bibfield
  {journal} {\bibinfo  {journal} {Arxiv Prepr.}\ }\textbf {\bibinfo {volume}
  {7}},\ \bibinfo {pages} {1504.02252} (\bibinfo {year} {2015})},\ \Eprint
  {http://arxiv.org/abs/1504.02252} {arXiv:1504.02252} \BibitemShut {NoStop}%
\bibitem [{\citenamefont {Zhang}\ \emph {et~al.}(2015)\citenamefont {Zhang},
  \citenamefont {Baker}, \citenamefont {Komineas},\ and\ \citenamefont
  {Hesjedal}}]{Zhang2015k}%
  \BibitemOpen
  \bibfield  {author} {\bibinfo {author} {\bibfnamefont {S.}~\bibnamefont
  {Zhang}}, \bibinfo {author} {\bibfnamefont {A.~A.}\ \bibnamefont {Baker}},
  \bibinfo {author} {\bibfnamefont {S.}~\bibnamefont {Komineas}}, \ and\
  \bibinfo {author} {\bibfnamefont {T.}~\bibnamefont {Hesjedal}},\ }\bibfield
  {title} {\enquote {\bibinfo {title} {{Topological computation based on direct
  magnetic logic communication}},}\ }\href {\doibase 10.1038/srep15773}
  {\bibfield  {journal} {\bibinfo  {journal} {Sci. Rep.}\ }\textbf {\bibinfo
  {volume} {5}},\ \bibinfo {pages} {15773} (\bibinfo {year}
  {2015})}\BibitemShut {NoStop}%
\bibitem [{\citenamefont {He}\ \emph {et~al.}(2016{\natexlab{b}})\citenamefont
  {He}, \citenamefont {Yin}, \citenamefont {Yu}, \citenamefont {Grutter},
  \citenamefont {Pan}, \citenamefont {Kou}, \citenamefont {Che}, \citenamefont
  {Yu}, \citenamefont {Nie}, \citenamefont {Zhang}, \citenamefont {Shao},
  \citenamefont {Murata}, \citenamefont {Zhu}, \citenamefont {Fan},
  \citenamefont {Han}, \citenamefont {Kirby},\ and\ \citenamefont
  {Wang}}]{He2016a}%
  \BibitemOpen
  \bibfield  {author} {\bibinfo {author} {\bibfnamefont {Q.~L.}\ \bibnamefont
  {He}}, \bibinfo {author} {\bibfnamefont {G.}~\bibnamefont {Yin}}, \bibinfo
  {author} {\bibfnamefont {L.}~\bibnamefont {Yu}}, \bibinfo {author}
  {\bibfnamefont {A.~J.}\ \bibnamefont {Grutter}}, \bibinfo {author}
  {\bibfnamefont {L.}~\bibnamefont {Pan}}, \bibinfo {author} {\bibfnamefont
  {X.}~\bibnamefont {Kou}}, \bibinfo {author} {\bibfnamefont {X.}~\bibnamefont
  {Che}}, \bibinfo {author} {\bibfnamefont {G.}~\bibnamefont {Yu}}, \bibinfo
  {author} {\bibfnamefont {T.}~\bibnamefont {Nie}}, \bibinfo {author}
  {\bibfnamefont {B.}~\bibnamefont {Zhang}}, \bibinfo {author} {\bibfnamefont
  {Q.}~\bibnamefont {Shao}}, \bibinfo {author} {\bibfnamefont {K.}~\bibnamefont
  {Murata}}, \bibinfo {author} {\bibfnamefont {X.}~\bibnamefont {Zhu}},
  \bibinfo {author} {\bibfnamefont {Y.}~\bibnamefont {Fan}}, \bibinfo {author}
  {\bibfnamefont {X.}~\bibnamefont {Han}}, \bibinfo {author} {\bibfnamefont
  {B.~J.}\ \bibnamefont {Kirby}}, \ and\ \bibinfo {author} {\bibfnamefont
  {K.~L.}\ \bibnamefont {Wang}},\ }\bibfield  {title} {\enquote {\bibinfo
  {title} {{Topological transitions induced by antiferromagnetism in a
  thin-film topological insulator}},}\ }\href {http://arxiv.org/abs/1612.01661}
  {\bibfield  {journal} {\bibinfo  {journal} {Arxiv Prepr.}\ } (\bibinfo {year}
  {2016}{\natexlab{b}})},\ \Eprint {http://arxiv.org/abs/1612.01661}
  {arXiv:1612.01661} \BibitemShut {NoStop}%
\bibitem [{\citenamefont {Ghosh}\ and\ \citenamefont
  {Manchon}(2017)}]{Ghosh2017}%
  \BibitemOpen
  \bibfield  {author} {\bibinfo {author} {\bibfnamefont {S.}~\bibnamefont
  {Ghosh}}\ and\ \bibinfo {author} {\bibfnamefont {A.}~\bibnamefont
  {Manchon}},\ }\bibfield  {title} {\enquote {\bibinfo {title} {{Spin-orbit
  torque in two-dimensional antiferromagnetic topological insulators}},}\
  }\href {\doibase 10.1103/PhysRevB.95.035422} {\bibfield  {journal} {\bibinfo
  {journal} {Phys. Rev. B}\ }\textbf {\bibinfo {volume} {95}},\ \bibinfo
  {pages} {035422} (\bibinfo {year} {2017})}\BibitemShut {NoStop}%
\bibitem [{\citenamefont {Kandala}\ \emph {et~al.}(2015)\citenamefont
  {Kandala}, \citenamefont {Richardella}, \citenamefont {Kempinger},
  \citenamefont {Liu},\ and\ \citenamefont {Samarth}}]{Kandala2015}%
  \BibitemOpen
  \bibfield  {author} {\bibinfo {author} {\bibfnamefont {A.}~\bibnamefont
  {Kandala}}, \bibinfo {author} {\bibfnamefont {A.}~\bibnamefont
  {Richardella}}, \bibinfo {author} {\bibfnamefont {S.}~\bibnamefont
  {Kempinger}}, \bibinfo {author} {\bibfnamefont {C.-X.}\ \bibnamefont {Liu}},
  \ and\ \bibinfo {author} {\bibfnamefont {N.}~\bibnamefont {Samarth}},\
  }\bibfield  {title} {\enquote {\bibinfo {title} {{Giant anisotropic
  magnetoresistance in a quantum anomalous Hall insulator}},}\ }\href {\doibase
  10.1038/ncomms8434} {\bibfield  {journal} {\bibinfo  {journal} {Nat.
  Commun.}\ }\textbf {\bibinfo {volume} {6}},\ \bibinfo {pages} {7434}
  (\bibinfo {year} {2015})},\ \Eprint {http://arxiv.org/abs/1503.0355}
  {arXiv:1503.0355} \BibitemShut {NoStop}%
\bibitem [{\citenamefont {Carbone}\ \emph {et~al.}(2016)\citenamefont
  {Carbone}, \citenamefont {Moras}, \citenamefont {Sheverdyaeva}, \citenamefont
  {Pacile}, \citenamefont {Papagno}, \citenamefont
  {Ferrari}, \citenamefont {Topwal}, \citenamefont {Vescovo}, \citenamefont
  {Bihlmayer}, \citenamefont {Freimuth}, \citenamefont {Mokrousov},\ and\
  \citenamefont {Bl{\"{u}}gel}}]{Carbone2016}%
  \BibitemOpen
  \bibfield  {author} {\bibinfo {author} {\bibfnamefont {C.}~\bibnamefont
  {Carbone}}, \bibinfo {author} {\bibfnamefont {P.}~\bibnamefont {Moras}},
  \bibinfo {author} {\bibfnamefont {P.~M.}\ \bibnamefont {Sheverdyaeva}},
  \bibinfo {author} {\bibfnamefont {D.}~\bibnamefont
  {Pacile}}, \bibinfo {author} {\bibfnamefont
  {M.}~\bibnamefont {Papagno}}, \bibinfo {author} {\bibfnamefont
  {L.}~\bibnamefont {Ferrari}}, \bibinfo {author} {\bibfnamefont
  {D.}~\bibnamefont {Topwal}}, \bibinfo {author} {\bibfnamefont
  {E.}~\bibnamefont {Vescovo}}, \bibinfo {author} {\bibfnamefont
  {G.}~\bibnamefont {Bihlmayer}}, \bibinfo {author} {\bibfnamefont
  {F.}~\bibnamefont {Freimuth}}, \bibinfo {author} {\bibfnamefont
  {Y.}~\bibnamefont {Mokrousov}}, \ and\ \bibinfo {author} {\bibfnamefont
  {S.}~\bibnamefont {Bl{\"{u}}gel}},\ }\bibfield  {title}
  {\enquote {\bibinfo {title} {{Asymmetric band gaps in a Rashba film
  system}},}\ }\href {\doibase 10.1103/PhysRevB.93.125409} {\bibfield
  {journal} {\bibinfo  {journal} {Phys. Rev. B - Condens. Matter Mater. Phys.}\
  }\textbf {\bibinfo {volume} {93}},\ \bibinfo {pages} {125409} (\bibinfo
  {year} {2016})}\BibitemShut {NoStop}%
\bibitem [{\citenamefont {Xue}(2011)}]{Xue2011}%
  \BibitemOpen
  \bibfield  {author} {\bibinfo {author} {\bibfnamefont {Q.-K.}\ \bibnamefont
  {Xue}},\ }\bibfield  {title} {\enquote {\bibinfo {title} {{Nanoelectronics: A
  topological twist for transistors}},}\ }\href {\doibase
  10.1038/nnano.2011.47} {\bibfield  {journal} {\bibinfo  {journal} {Nat.
  Nanotechnol.}\ }\textbf {\bibinfo {volume} {6}},\ \bibinfo {pages} {197--198}
  (\bibinfo {year} {2011})}\BibitemShut {NoStop}%
\bibitem [{\citenamefont {Liu}\ \emph {et~al.}(2013)\citenamefont {Liu},
  \citenamefont {Hsieh}, \citenamefont {Wei}, \citenamefont {Duan},
  \citenamefont {Moodera},\ and\ \citenamefont {Fu}}]{Liu2013d}%
  \BibitemOpen
  \bibfield  {author} {\bibinfo {author} {\bibfnamefont {J.}~\bibnamefont
  {Liu}}, \bibinfo {author} {\bibfnamefont {T.~H.}\ \bibnamefont {Hsieh}},
  \bibinfo {author} {\bibfnamefont {P.}~\bibnamefont {Wei}}, \bibinfo {author}
  {\bibfnamefont {W.}~\bibnamefont {Duan}}, \bibinfo {author} {\bibfnamefont
  {J.}~\bibnamefont {Moodera}}, \ and\ \bibinfo {author} {\bibfnamefont
  {L.}~\bibnamefont {Fu}},\ }\bibfield  {title} {\enquote {\bibinfo {title}
  {{Spin-filtered Edge States with an Electrically Tunable Gap in a
  Two-Dimensional Topological Crystalline Insulator}},}\ }\href {\doibase
  10.1038/nmat3828} {\bibfield  {journal} {\bibinfo  {journal} {Nat. Mater.}\
  }\textbf {\bibinfo {volume} {13}},\ \bibinfo {pages} {178--183} (\bibinfo
  {year} {2013})},\ \Eprint {http://arxiv.org/abs/1310.1044} {arXiv:1310.1044}
  \BibitemShut {NoStop}%
\bibitem [{\citenamefont {Fujita}(2016)}]{Fujita2016}%
  \BibitemOpen
  \bibfield  {author} {\bibinfo {author} {\bibfnamefont {H.}~\bibnamefont
  {Fujita}},\ }\bibfield  {title} {\enquote {\bibinfo {title} {{Field-free,
  spin-current control of magnetization in non-collinear chiral
  antiferromagnets}},}\ }\href {\doibase 10.1002/pssr.201600360} {\bibfield
  {journal} {\bibinfo  {journal} {Phys. status solidi - Rapid Res. Lett.}\ }
  (\bibinfo {year} {2016}),\ 10.1002/pssr.201600360},\ \Eprint
  {http://arxiv.org/abs/1610.07615} {arXiv:1610.07615} \BibitemShut {NoStop}%
\bibitem [{\citenamefont {Feng}\ \emph {et~al.}(2015)\citenamefont {Feng},
  \citenamefont {Guo}, \citenamefont {Zhou}, \citenamefont {Yao},\ and\
  \citenamefont {Niu}}]{Feng2015}%
  \BibitemOpen
  \bibfield  {author} {\bibinfo {author} {\bibfnamefont {W.}~\bibnamefont
  {Feng}}, \bibinfo {author} {\bibfnamefont {G.-Y.}\ \bibnamefont {Guo}},
  \bibinfo {author} {\bibfnamefont {J.}~\bibnamefont {Zhou}}, \bibinfo {author}
  {\bibfnamefont {Y.}~\bibnamefont {Yao}}, \ and\ \bibinfo {author}
  {\bibfnamefont {Q.}~\bibnamefont {Niu}},\ }\bibfield  {title} {\enquote
  {\bibinfo {title} {{Large magneto-optical Kerr effect in noncollinear
  antiferromagnets Mn 3 X ( X = Rh , Ir , Pt )}},}\ }\href {\doibase
  10.1103/PhysRevB.92.144426} {\bibfield  {journal} {\bibinfo  {journal} {Phys.
  Rev. B}\ }\textbf {\bibinfo {volume} {92}},\ \bibinfo {pages} {144426}
  (\bibinfo {year} {2015})}\BibitemShut {NoStop}%
\bibitem [{\citenamefont {Thouless}(1998)}]{thouless1998topological}%
  \BibitemOpen
  \bibfield  {author} {\bibinfo {author} {\bibfnamefont {D.}~\bibnamefont
  {Thouless}},\ }\href {https://books.google.cz/books?id=6XC\_PBEXnAEC} {\emph
  {\bibinfo {title} {Topological Quantum Numbers in Nonrelativistic Physics}}}\
  (\bibinfo  {publisher} {World Scientific},\ \bibinfo {year}
  {1998})\BibitemShut {NoStop}%
\end{thebibliography}
\end{document}